\documentclass[pageno]{jpaper}

\usepackage{amsmath,amsfonts}
\usepackage{algorithm}
\usepackage{algorithmic}

\floatname{algorithm}{Algorithm}

\newcommand{\Name}{MP-Rec}
\newcommand{\Cache}{MP-Cache}
\newcommand{\CacheOne}{\Cache$_{encoder}$}
\newcommand{\CacheTwo}{\Cache$_{decoder}$}
\newcommand{\newMetric}{$throughput_{correct\_predictions}$}

\usepackage[normalem]{ulem}

\begin{document}

\title{\Name: Hardware-Software Co-Design to Enable Multi-Path Recommendation}
\author{
Samuel Hsia$^{1,2}$, Udit Gupta$^{1}$, Bilge Acun$^{1}$, Newsha Ardalani$^{1}$, Pan Zhong$^{1}$, \\
Gu-Yeon Wei$^{2}$, David Brooks$^{1,2}$, Carole-Jean Wu$^{1}$\\ \\
$^{1}$Meta AI, $^{2}$Harvard University\\ \\
shsia@g.harvard.edu, carolejeanwu@meta.com
}

\date{}
\maketitle

\pagenumbering{arabic}

\begin{abstract}
Deep learning recommendation systems serve personalized content under diverse tail-latency targets and input-query loads.
In order to do so, state-of-the-art recommendation models rely on terabyte-scale embedding tables to learn user preferences over large bodies of contents.
The reliance on a \textit{fixed} embedding representation of embedding tables not only imposes significant memory capacity and bandwidth requirements but also limits the scope of compatible system solutions.
This paper challenges the assumption of fixed embedding representations by showing how synergies between embedding representations and hardware platforms can lead to improvements in both algorithmic- and system performance.
Based on our characterization of various embedding representations, we propose a \textit{hybrid} embedding representation that achieves higher quality embeddings at the cost of increased memory and compute requirements.
To address the system performance challenges of the \textit{hybrid} representation, we propose \textit{\Name} --- a co-design technique that exploits heterogeneity and dynamic selection of embedding representations and underlying hardware platforms.

On real system hardware, we demonstrate how matching custom accelerators, i.e., GPUs, TPUs, and IPUs, with compatible embedding representations can lead to $16.65\times$ performance speedup. 
Additionally, in query-serving scenarios, \Name~achieves 2.49$\times$ and 3.76$\times$ higher correct prediction throughput and 0.19\% and 0.22\% better model quality on a CPU-GPU system for the Kaggle and Terabyte datasets, respectively.
\end{abstract}

\section{Introduction}~\label{sec:intro}
Deep learning (DL) recommendation models support a wide variety of applications, such as search~\cite{anil2022factoryfloor, cheng2016wnd, jouppi2017tpu, zhao2019mtwnd}, social media~\cite{acun2020understanding, gupta2020architectural, hazelwood2018applied, yiSysml18}, e-commerce~\cite{zhou2019dien, zhou2018din}, and entertainment~\cite{he2017ncf}.
Due to its overarching impact, neural recommendation has become a dominant source of compute cycles in large-scale AI infrastructures.
In 2019, recommendation use cases contributed to 79\% of the overall AI inference cycles at Meta, making it one of the most resource-demanding DL use cases at the data center scale~\cite{gupta2020architectural, hazelwood2018applied}.

A critical component of state-of-the-art recommendation models is the embedding table~\cite{cheng2016wnd, he2017ncf, lui2021capacity, naumov2019dlrm, yin2021ttrec, zhao2019mtwnd, zhou2019dien, zhou2018din}.
Information, such as user preferences and content understanding, is represented as individual vectors within these embedding tables.
To support increasingly complex applications and user preference models, embedding table sizes have grown super-linearly into the terabyte-scale~\cite{wu2022sustainableai}.
As a result, a plethora of system- and hardware-level solutions for neural recommendation have focused on addressing the memory capacity and bandwidth challenges of large-scale embedding tables~\cite{ke2020recnmp, kwon2019tensordimm, wilkening2021recssd, jiang2021fleetrec, lui2021capacity, anderson2021fbinference, deng2021lowprecision, khudia2021fbgemm}.

\begin{figure}[t!]
    \centering
    \includegraphics[width=\linewidth]{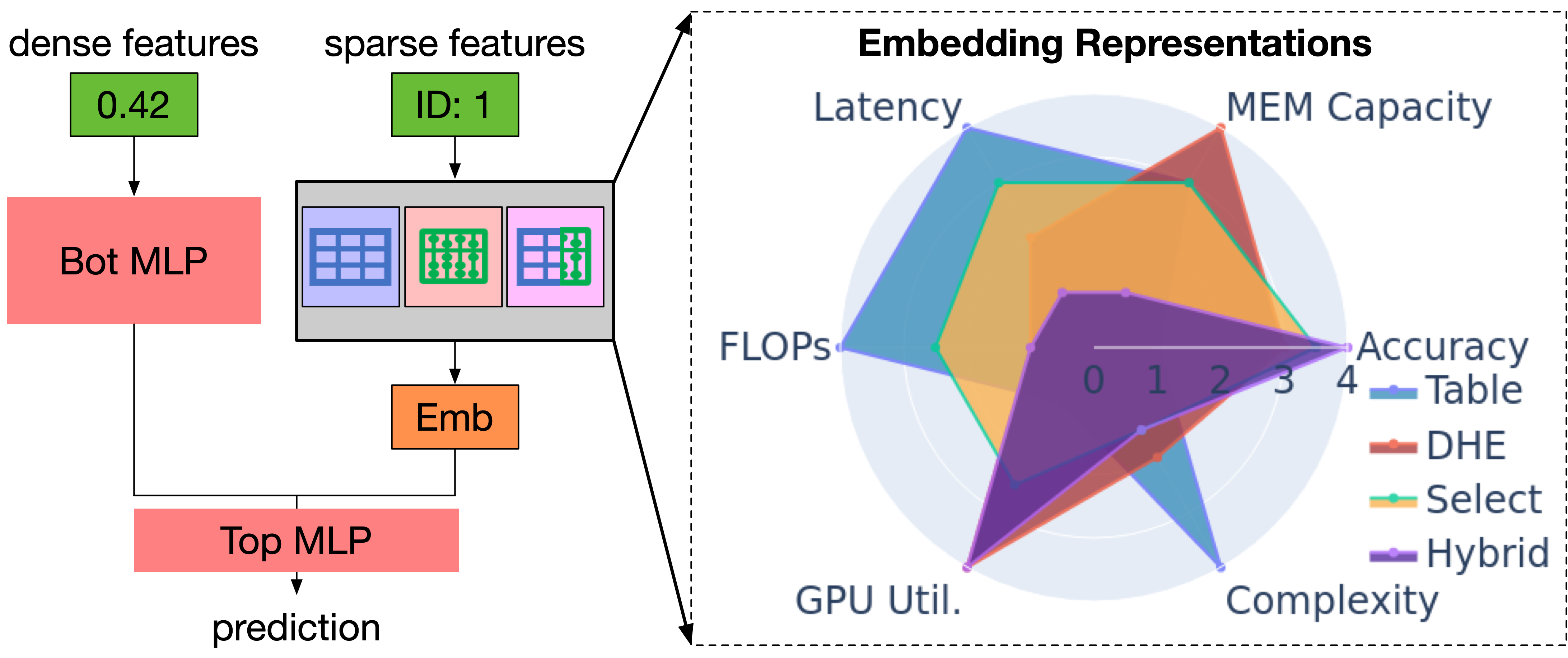}
    \caption{Compared to prior work, \Name~explores the embedding access design space through different \textit{embedding representations}.}
    \label{fig:motivation}
\end{figure}

However, there is additional room for algorithmic and system performance improvements if we go beyond exclusively using tables as \textit{embedding representation}.
Recent proposals examine alternative embedding representations use GEMM-heavy compute stacks to dynamically generate embedding vectors~\cite{yin2021ttrec, kang2021dhe}.
While these representations significantly reduce memory capacity requirements, the techniques introduce orders of magnitude higher FLOPs.

Based on the detailed design space characterization for embedding representations (Figure \ref{fig:motivation}; Section \ref{sec:characterization}), we identify significant performance improvement potential when utilizing custom accelerators for compatible representations.   
To demonstrate the impact of representation-hardware compatibility, we perform real-system evaluations on a wide range of hardware, including CPUs and GPUs, as well as custom AI accelerators such as Google Tensor Processing Units (TPUs)~\cite{jouppi2021tpu, jouppi2020tpu, jouppi2017tpu} and Graphcore Intelligence Processing Units (IPUs)~\cite{moor2020ipu, jia2019ipu} at core-, chip-, board, and pod-level configurations, demonstrating up to 16.65$\times$ performance speedup.
Ultimately, \textit{there is no one-size-fits-all static solution} as representation requirements and hardware capabilities vary.

In this work, we propose a new \textit{hybrid} embedding representation for neural recommendation tasks.
While prior representations focus exclusively on either memory- or compute-based execution paths, the \textit{hybrid} representation leverages these contrasting execution patterns to increase learning capacity and produce higher quality embeddings.
Our evaluation results show that the \textit{hybrid} representation increases model quality by 0.19\%, 0.22\%, and 0.014\% on the open-source Kaggle~\cite{criteokaggle}, Terabyte~\cite{criteoterabyte}, and internal use-cases, respectively.
Due to its increased complexity, the \textit{hybrid} representation requires even higher capacity and memory requirements.

Taking a step further, to support the resource-intensive \textit{hybrid} representation and dynamic mapping of heterogeneous accelerators and representations, we propose Multi-Path Recommendation (\Name) --- a dynamic representation-hardware co-design technique to maximize throughput of correct recommendations while meeting tail latency requirements.
Depending on memory capacities of AI inference systems, \Name~first generates accuracy-optimal representation-hardware mappings based on the unique properties of different embedding representations (Figure \ref{fig:motivation} (right)), forming multiple potential embedding execution paths.
At runtime, depending on input query sizes and application-specific performance targets, \Name~activates embedding path(s) by scheduling queries onto available representation-hardware configurations to jointly maximize prediction quality and throughput.
To further speed up \Name, we introduce \Cache~to exploit novel caching opportunities introduced by the intermediate results of the new embedding representations.
\Cache~targets both data locality and value similarity of embedding accesses to make computationally-expensive representations viable.

We evaluate the performance of \Name~on a real CPU-GPU platform, demonstrating 2.49$\times$ and 3.76$\times$ higher throughput of correct predictions on Kaggle~\cite{criteokaggle} and Terabyte~\cite{criteoterabyte}, respectively, over the CPU baseline.
In addition, when we integrate IPUs into \Name~for query serving, we observe a significant throughput improvement potential of 34.24$\times$ that can be unlocked with future software support.
For the constant throughput scenarios at strict SLA latency targets, \Name~reduces SLA latency target violations by 27.59\% compared to exclusively using the baseline embedding table-based recommendation models on CPUs.
Overall, \Name~showcases the massive potential of algorithmic- and system-performance improvements when we integrate embedding representation design into the system design space for neural recommendation.
Finally, these performance improvements can be further enhanced by custom AI accelerators that benefit from representation-level synergies.

The main contributions of this work are as follows:
\begin{itemize}
    \item We propose a new \textit{hybrid} embedding representation that increases learning capacity and produces higher quality embeddings at the cost of increased compute and memory requirements.
    The \textit{hybrid} embedding representation demonstrates measurable improvements in model quality.
    \item We implement and characterize different embedding representations using state-of-the-art custom AI accelerators (i.e., TPUs and IPUs) at core-, chip-, board-, and pod-level configurations.
    We identify key system challenges of adapting specific representations to accelerators, highlighting distinct accelerator-specific advantages: TPUs for embedding tables, IPUs for compact compute-stacks, and GPUs for energy-efficient execution of large-capacity models (Section~\ref{ssec:char_accelerators}).
    \item We propose \Name~--- a dynamic representation-hardware co-design technique for deep learning recommendation inference.
    \Name~mitigates the performance and accuracy degradations from static representation-hardware mappings.
    We augment \Name~with a two-tier cache design (\Cache) to exploit unique caching opportunities found in compute-based representations.
\end{itemize}
\section{Background: Embedding Representations}~\label{sec:background}

\begin{figure}[t!]
    \centering
    \includegraphics[width=\linewidth]{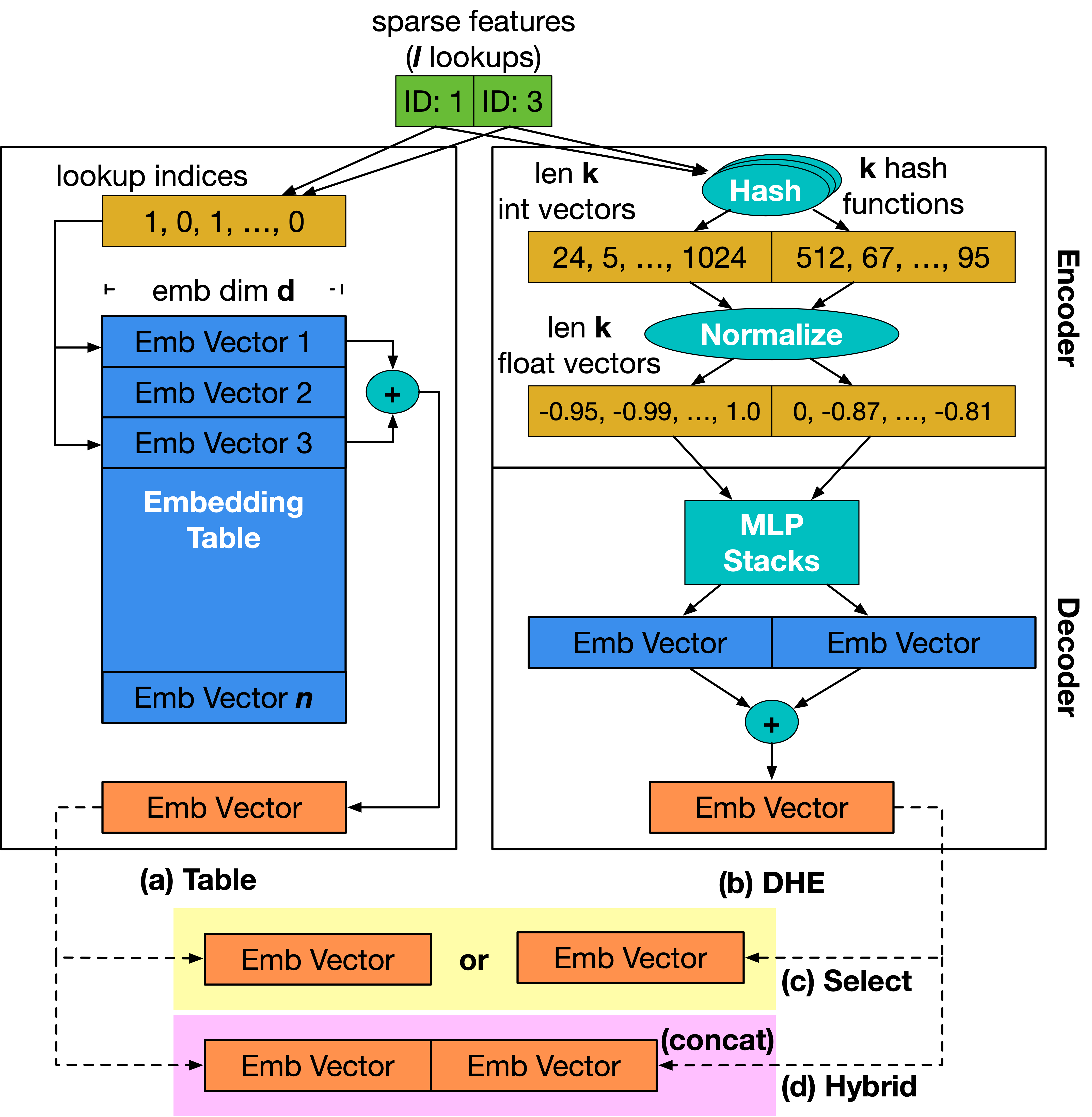}
    \caption{
    \underline{(a)} \textit{Table} representation stores learned embeddings while \underline{(b)} \textit{DHE} dynamically generates embeddings via encoder-decoder stacks. 
    We introduce \underline{(c)} \textit{select} representation that selects either (a) or (b) at table-level granularity and \underline{(d)} \textit{hybrid} representation that leverages both (a) and (b) for highly accurate embeddings.
    }
    \label{fig:representations}
\end{figure}

Embeddings are a performance-critical component of neural recommendation models.
In order to be processed by recommendation models, sparse feature IDs must first be transformed into dense embedding vectors.
This transformation process -- \textit{embedding access} -- can be realized through different embedding representations.

We start with two distinct classes of embedding representations: \textit{storage} and \textit{generation}.
While storing learned embedding vectors as tables is a widely adopted approach, it introduces significant memory capacity and bandwidth requirements~\cite{gholami2021memorywall, gupta2020deeprecsys, hsia2020cross, ke2020recnmp, kwon2019tensordimm, lui2021capacity, wilkening2021recssd} (Section \ref{ssec:bgd_tbl}). 
On the other hand, generating embeddings with compute-heavy encoder-decoder stacks trades off these memory system requirements with compute demand (Section \ref{ssec:bgd_dhe}).
To leverage these contrasting qualities, we introduce new embedding representations -- \textit{select} and \textit{hybrid} -- that leverage complementary system resources to generate embeddings from table and DHE representations (Section \ref{ssec:bgd_newreps}).


\subsection{Storing Embeddings: Embedding Tables}~\label{ssec:bgd_tbl}
Figure \ref{fig:representations} (a) depicts the embedding access mechanism of a typical neural recommendation model.
Sparse feature lookup IDs are converted into multi-hot encoded vectors, which are then used as indices into embedding tables. 
While the table representation is used by many state-of-the-art neural recommender systems for its simplicity and relatively high accuracy, it has significant memory bandwidth and capacity requirements.

\subsection{Generating Embeddings: DHE}~\label{ssec:bgd_dhe}
Alternatively, embedding vectors can be dynamically generated from compute stacks.
Examples include Tensor Train Compression (TT-Rec)~\cite{yin2021ttrec} and Deep Hash Embedding (DHE)~\cite{kang2021dhe}.
In this work, we focus on DHE (Figure \ref{fig:representations}(b)) over TT-Rec due to the flexibility in tuning DHE's encoder-decoder stacks.
DHE execution is separated into two phases: the encoder stack generates intermediate values from sparse ID inputs (upper block) and the decoder stack generates dense embedding vectors from the intermediate values (lower block).
More specifically, the functionalities of the encoder and decoder stacks are as follows:
\begin{itemize}
    \item \textbf{(Encoder Stack):}
    First apply $k$ parallel, unique encoder hash functions on input sparse IDs.
    Instead of being used as lookup indices, these updated IDs are then applied with normalization functions to create intermediate dense features.
    \item \textbf{(Decoder Stack):}
    The intermediate dense features then pass through decoder MLP stacks to generate final embedding vectors for downstream stacks.
\end{itemize}
While embedding tables and DHE both produce dense embedding vectors from sparse IDs, the required algorithmic steps and compute resources are fundamentally different. 
For embedding tables, individual embedding vectors are stored after training and accessed during inference.

To learn valuable correlations from the ever-increasing data volume, the number of entries per embedding table have bloated to millions, leading to aggregate memory capacity requirements in the terabytes~\cite{lui2021capacity, mudigere2021zionex}.
In DHE, encoder-decoder stacks are first trained offline.
During inference, embedding vectors are dynamically generated running sparse IDs through trained DHE stacks.
Encoder hash functions and decoder MLPs contribute higher FLOPs, shifting the system bottleneck from memory capacity to computation.

\subsection{Novel Representations: Select \& Hybrid}~\label{ssec:bgd_newreps}
In Figure \ref{fig:representations}(c), we present the proposed \textit{select embedding representation}, where we select either embedding table or DHE representation at the feature-level (i.e., table-level) granularity.
With the \textit{select} embedding representation, recommendation model designers can make the aforementioned memory-compute tradeoffs for each sparse feature.
In Figure \ref{fig:representations}(d), we capitalize upon the dichotomy of embedding tables and DHE by proposing a \textit{table-compute hybrid representation}.
In this proposed \textit{hybrid} representation, sparse IDs are used to \textit{both} access embedding tables and dynamically generate embedding vectors.
The resulting embeddings from both mechanisms are then concatenated.
The embedding tables and decoder MLP stacks are trained together.

Our newly proposed \textit{hybrid} representation is based on two key observations. 
First, embeddings learned from tables and DHE have different semantics.
Generated embeddings from DHE can achieve higher model quality for some CTR predictions tasks~\cite{criteokaggle, reddi2020mlperfinference, criteoterabyte}, as demonstrated in Section \ref{ssec:char_accuracy}.
Second, embedding tables and DHE compute stacks stress independent system resources.
The \textit{hybrid} representation is unique in its ability to fully utilize both memory and compute resources of an underlying system for higher recommendation quality.
\section{Design Space Exploration for Sparse Feature Representation}~\label{sec:characterization}
In this section, we characterize embedding table, DHE, \textit{select}, and \textit{hybrid} representations across the important dimensions of model accuracy (Section~\ref{ssec:char_accuracy}), model capacity (Section~\ref{ssec:char_capacity}), execution latency (Section~\ref{ssec:char_latency}), and accelerator compatibility (Section~\ref{ssec:char_accelerators}).
We show that this previously unexplored design space offers not only noticeable accuracy improvements but also hardware-specific optimization opportunities.

\begin{figure*}[t!]
    \centering
    \includegraphics[width=\linewidth]{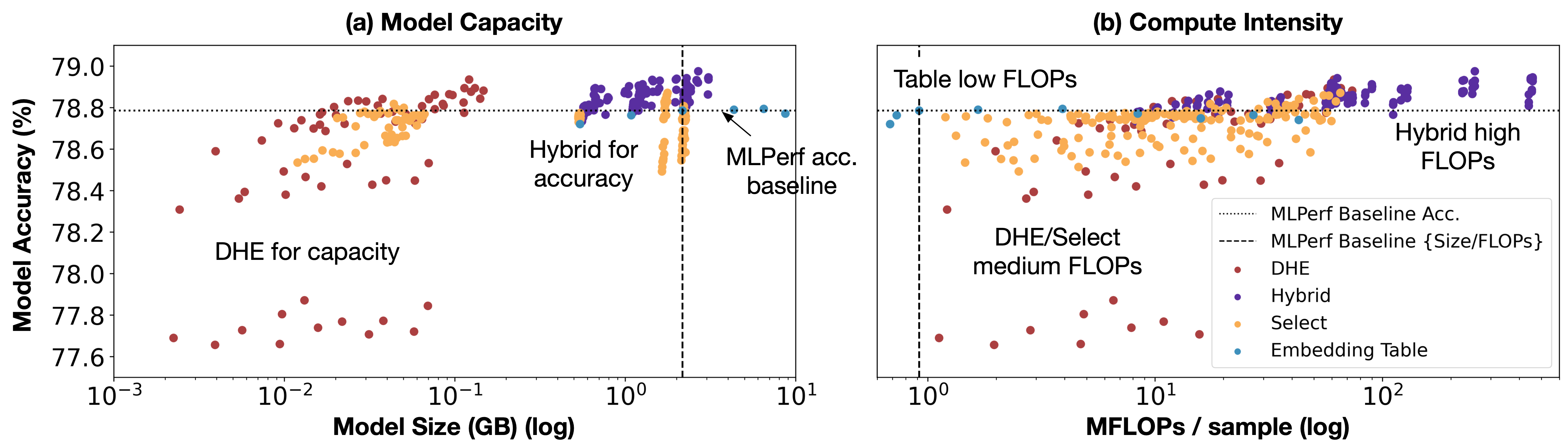}
    \caption{\textit{DHE} representation saves memory capacity, \textit{hybrid} representation enables optimal accuracies, and \textit{table} representation has faster latency from less FLOPs. Evaluation is on the Criteo Kaggle data set.} 
    \label{fig:characterization}
\end{figure*}

Figure \ref{fig:characterization} provides an overview for the design space trade-offs of the four embedding representations along model accuracy (y-axis), capacity, and FLOPs (x-axis for (a) and (b), respectively).

\subsection{Accuracy: Tuning DHE Parameters}~\label{ssec:char_accuracy}
\textit{Hybrid} representation (violet points) configurations, which use both table and compute stacks, achieve the highest accuracies.
Figure \ref{fig:characterization} illustrates that the most accurate \textit{hybrid} representation configurations achieve $0.19\%$ and $0.22\%$ accuracy improvements over the embedding table baselines on Kaggle and Terabyte, respectively.
Note that for many recommendation use cases, accuracy improvements $>0.1\%$ are considered significant.~\cite{anil2022factoryfloor, wang2018alibaba}.

We tune the hyperparameters of a DHE stack
to realize these improved accuracies.
Each point in Figure \ref{fig:characterization} corresponds to a model with unique hyperparameters. 
For embedding tables, we vary embedding dimension.
For DHE stacks, we vary the number of parallel encoder hash functions $k$, the decoder MLP width $d_{NN}$, and decoder MLP height $h$.
We also vary the shape of FC layers for each decoder MLP stack.

\begin{figure}[t!]
    \centering
    \includegraphics[width=\linewidth]{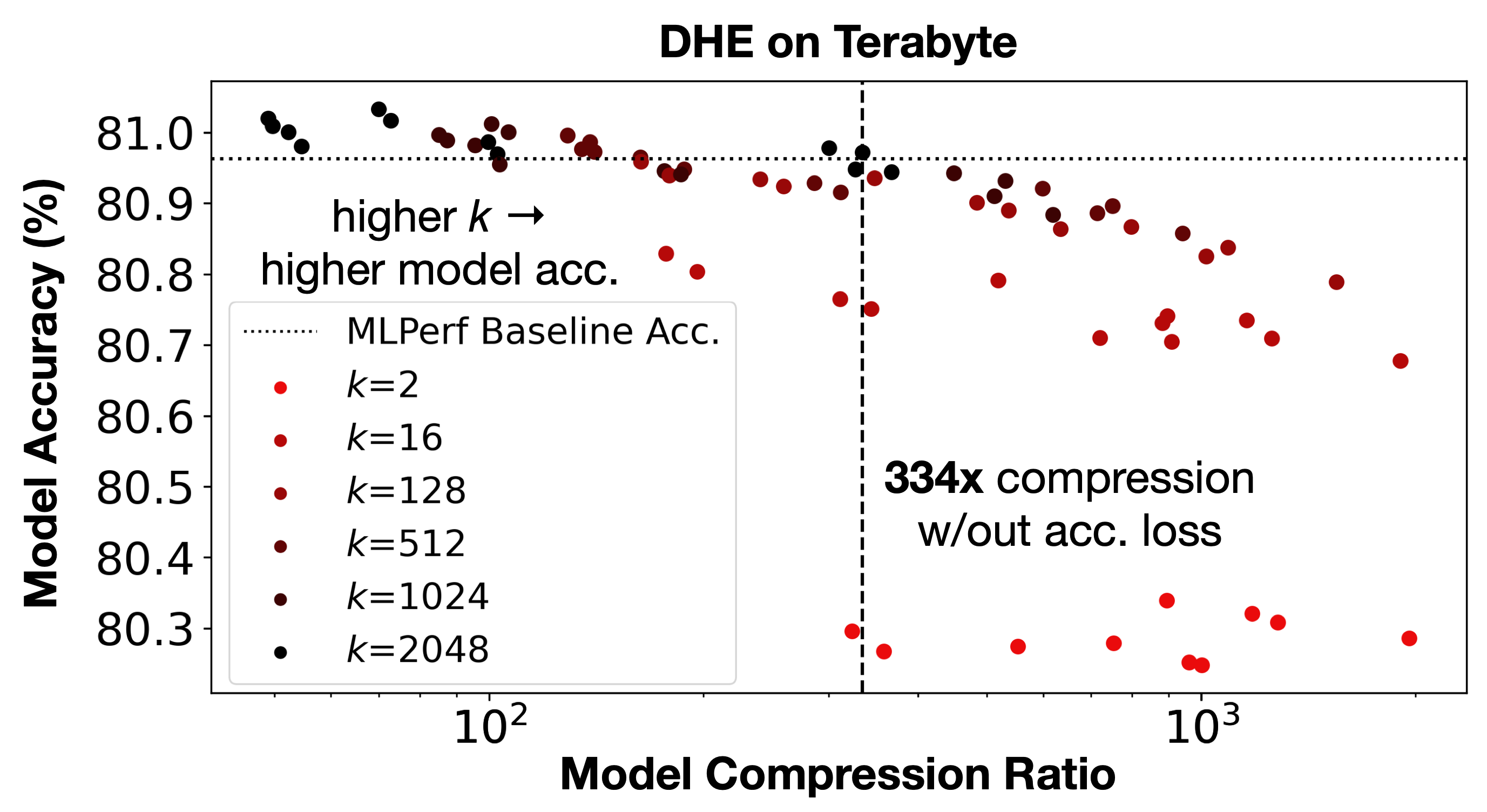}
    \caption{DHE compute stacks can be tuned to improve either model \textit{accuracy} or \textit{compression} ratio.}
    \label{fig:accuracy}
\end{figure}

\begin{figure}[t!]
    \centering
    \includegraphics[width=\linewidth]{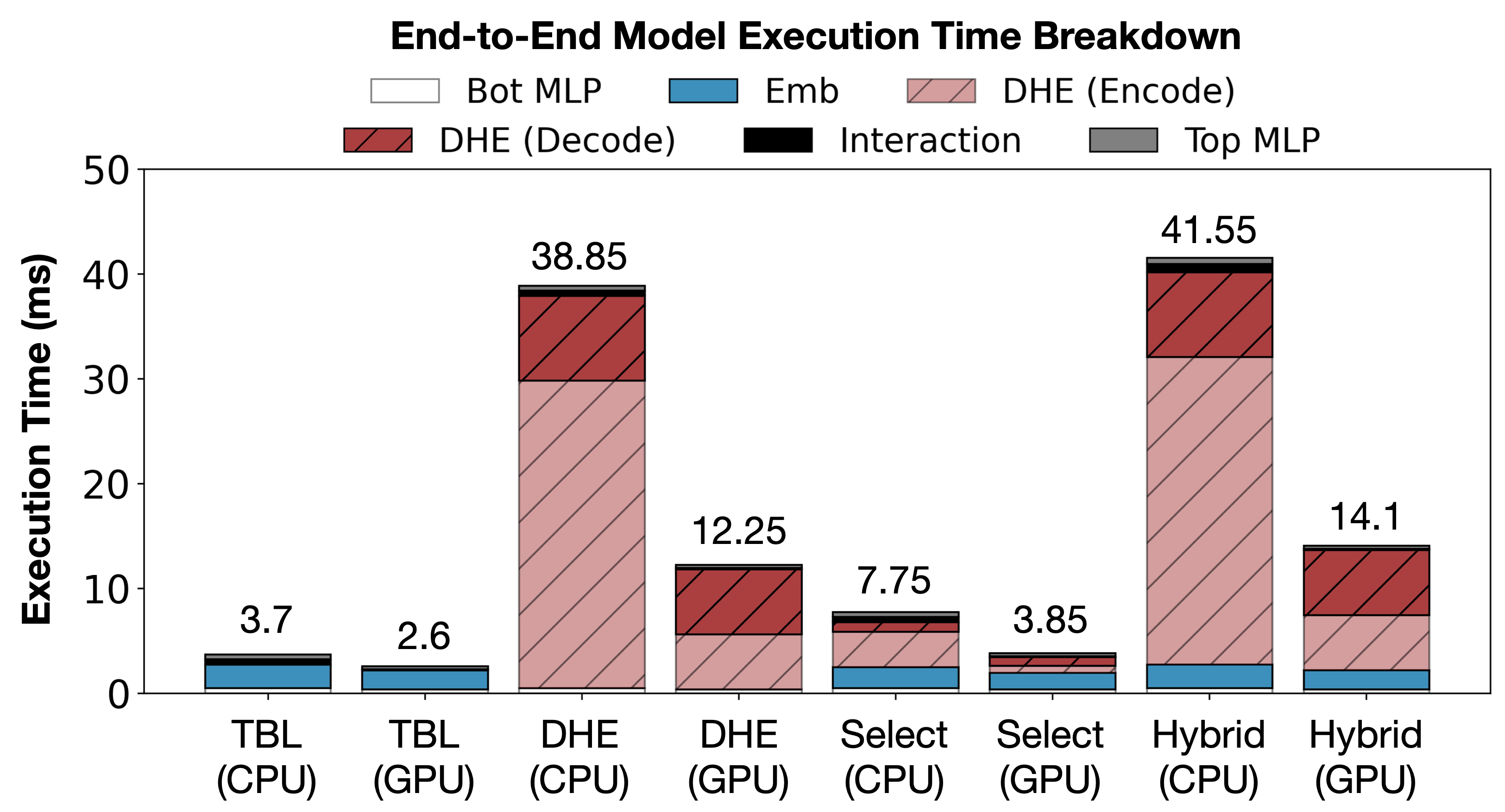}
    \caption{\textit{Operator breakdown} of Table, DHE, \textit{select}, and \textit{hybrid} execution on CPUs and GPUs. \textit{Hybrid} shows worst performance in terms of latency while \textit{select} offers a compromise between Table and DHE.}
    \label{fig:latency}
\end{figure}

\begin{figure}[t!]
    \centering
    \includegraphics[width=\linewidth]{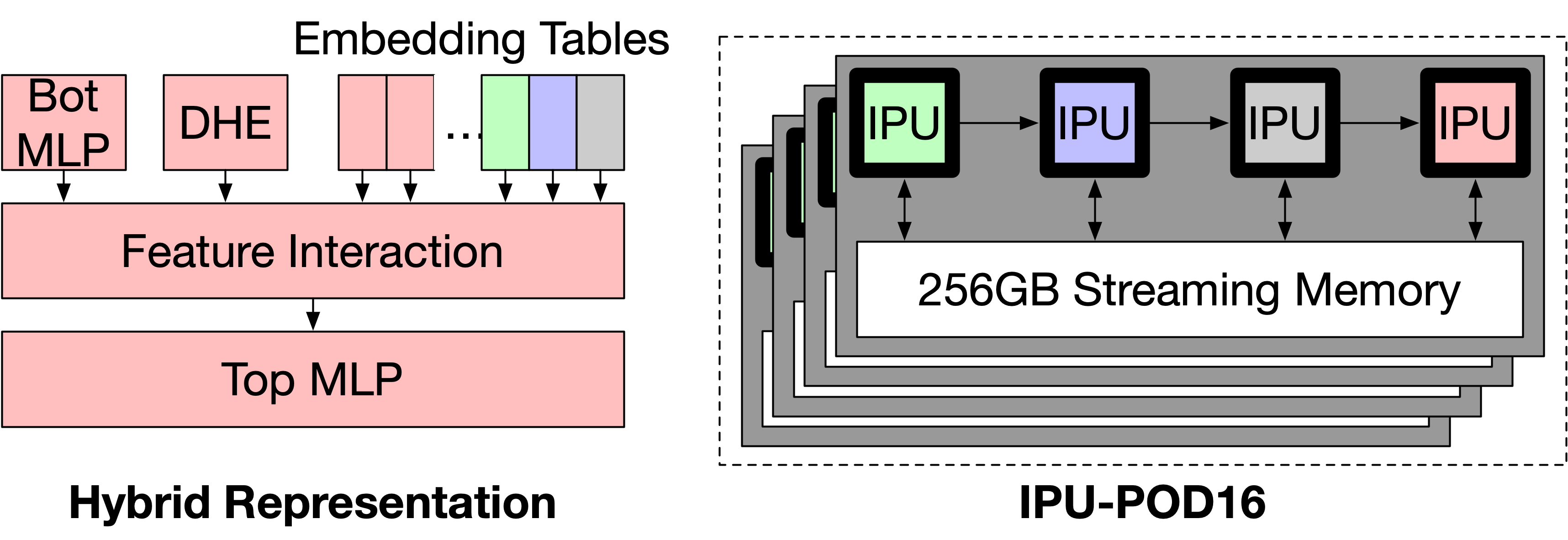}
    \caption{
    \textit{Hybrid} representation deployment strategies for pod-scale IPUs.
    Single chip execution requires offloading three largest embedding tables (green, blue, and gray shade) to DRAM while pod-level execution duplicates board-level parallelism strategy four times for data parallelism.}
    \label{fig:ipu_parallelism}
\end{figure}

\begin{figure*}[t!]
    \centering
    \includegraphics[width=\linewidth]{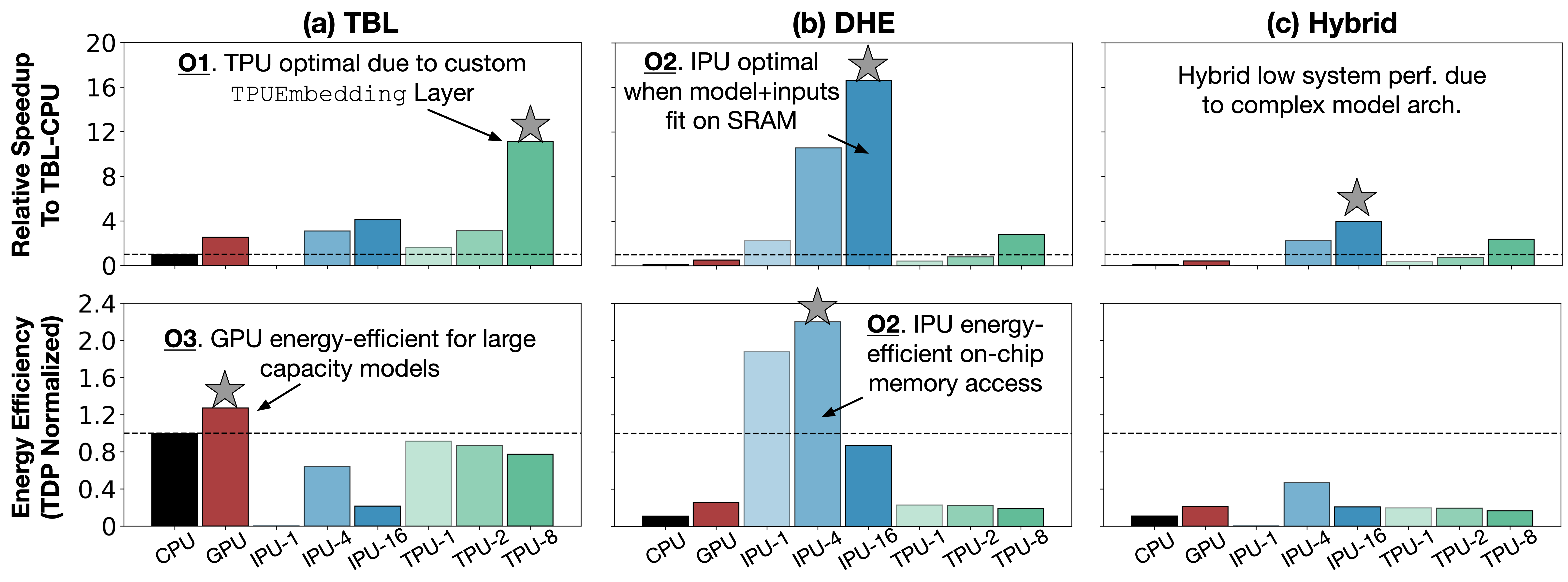}
    \caption{\textit{Table, DHE, and \textit{hybrid} embedding representations} evaluated across different custom accelerators. TPUv3s see speedups from TPU-optimized \texttt{TPUEmbedding}s while Graphcore IPUs offer optimal performance when the model and activations fit within its 900 MB SRAM per-chip scratchpad.
    CPU: Broadwell Xeon; GPU: V100; IPU-1: 1-chip GC200; TPU-1: 1-core TPUv3 (TPU has 2 cores/chip).}
    \label{fig:accelerators}
\end{figure*}

Figure \ref{fig:accuracy} depicts the compression ratio -- relative to a 12.59 GB embedding table baseline model -- (x-axis) and model accuracy (y-axis) for different DHE configurations.
The color of each point denotes the number of hash functions used ($k$).
We see that, as $k$ increases from 2 to 2048 (i.e., color progression from red to black), model accuracy increases.
Thus, $k$ is an important factor in determining achievable model accuracy.
In contrast, for a given $k$, varying the decoder MLP size and shape had a relatively insignifcant effect on model accuracy.
This can be observed from how points with the same color (i.e., same $k$ different ($d_{NN}$, $h$)) have relatively similar model accuracies.
A similar trend is observed for \textit{select} and \textit{hybrid} representations as well.

\subsection{Capacity: Enabling Memory-Constrained Neural Recommendation}~\label{ssec:char_capacity}
In Figure \ref{fig:characterization} (a), we observe that DHE configurations (red points) have model capacities $10\sim1000\times$ smaller than gigabyte-scale baseline embedding table configurations (blue points).
Through DHE, we are able to construct a recommendation model that is $334\times$ smaller in model capacity than the MLPerf baseline which uses embedding tables -- without any accuracy degradation (i.e., horizontal dotted MLPerf accuracy baseline)~\cite{reddi2020mlperfinference, criteoterabyte}.
DHE accomplishes this by having a shared set of encoder-decoder parameters for generating embeddings as opposed to storing user- and item-specific embeddings.
With these compression ratios, recommendation models can be compressed by orders of magnitude, from GBs to MBs, and deployed on a wider range of hardware platforms and use-cases.

\subsection{Latency: Operator Breakdowns}
~\label{ssec:char_latency}

\begin{figure*}[t!]
    \centering
    \includegraphics[width=\linewidth]{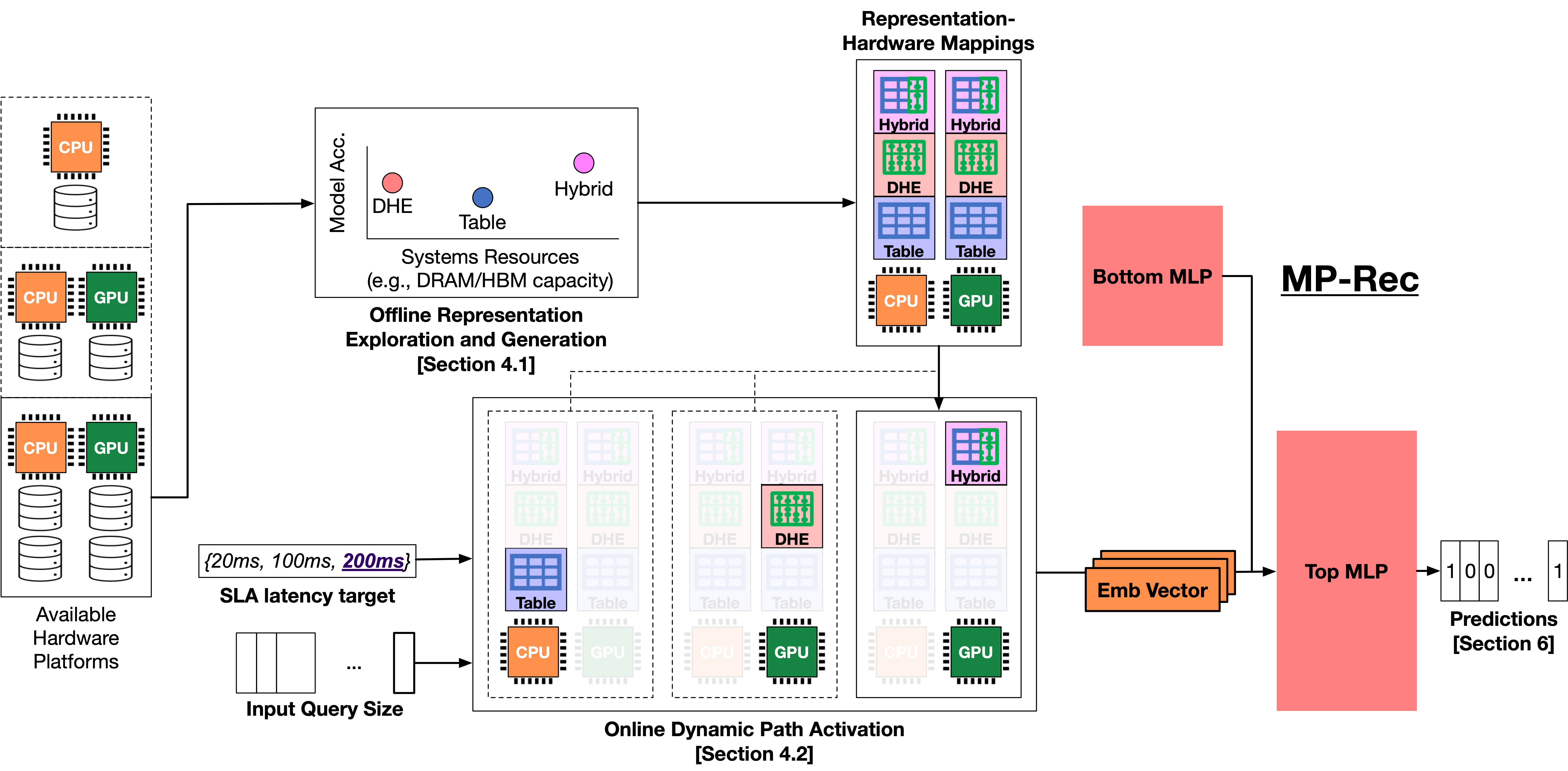}
    \caption{
    \textit{\Name}~has two stages: offline mapping exploration and online query scheduler.
    In the offline phase, \Name~considers algorithmic and systems-level exploration constraints to generate optimal representation-hardware configurations.
    In the online phase, \Name~dynamically schedules queries onto available representation-HW execution paths based on query-level information to maximize for amount of high quality recommendation.}
    \label{fig:design}
\end{figure*}

\begin{figure}[t!]
    \centering
    \includegraphics[width=\linewidth]{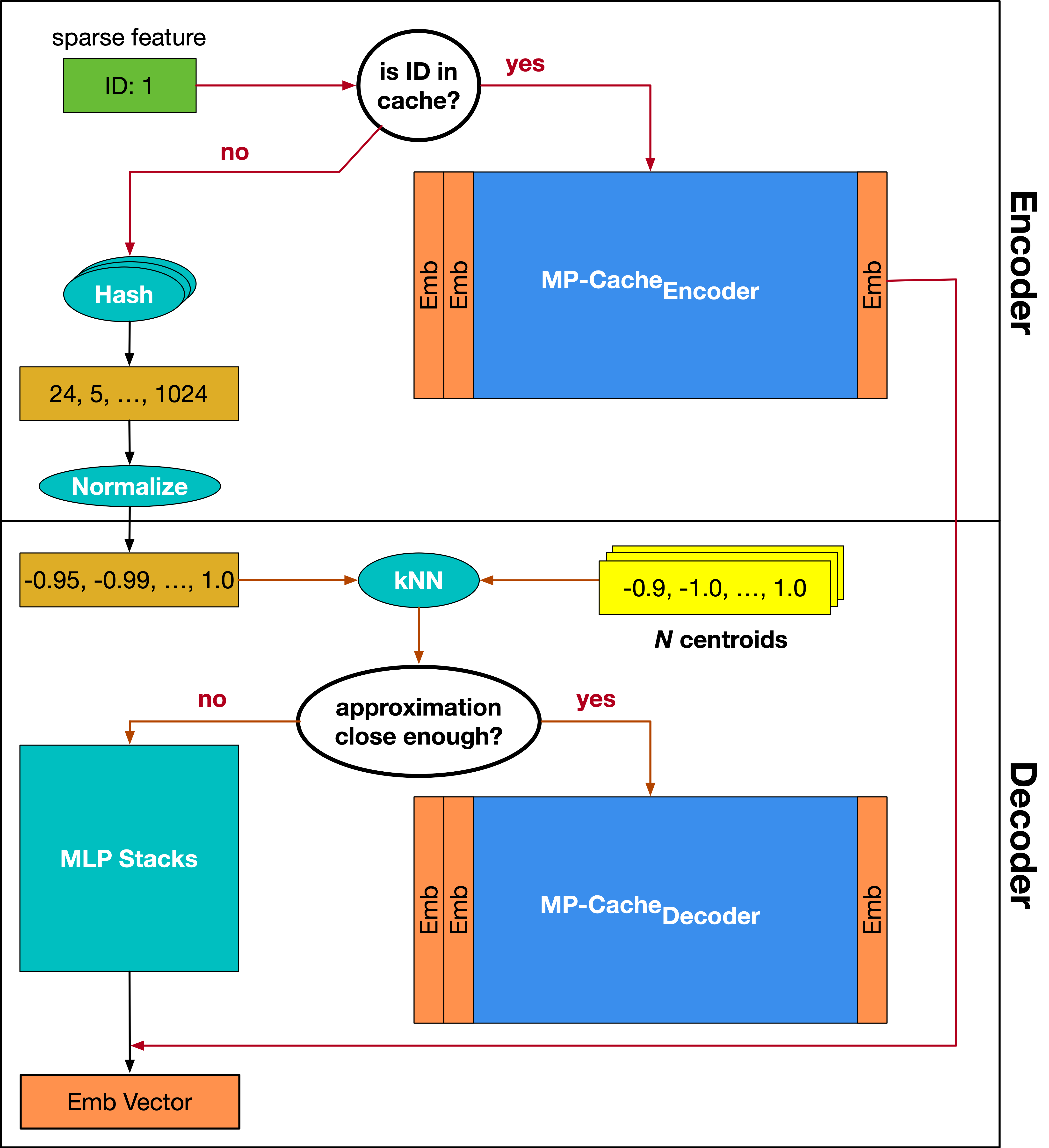}
    \caption{
    \textit{\Cache}~is comprised of two cascading stages.
    \CacheOne~exploits access frequency of sparse IDs while \CacheTwo~exploits value similarity between intermediate results.}
    \label{fig:cache}
\end{figure}

While compute-based representations, such as DHE and \textit{hybrid}, can improve model accuracies, the encoder-decoder stacks introduce FLOPs that lead to execution slowdowns.
Figure \ref{fig:characterization} (b) shows how models that use DHE and \textit{hybrid} have $10\sim100\times$ more FLOPs than those relying on embedding tables.

Figure \ref{fig:latency} shows operator breakdown of different representations for CPUs and GPUs.
For DHE, we see $10.5\times$ and $4.7\times$ slowdown on CPUs and GPUs, respectively.
DHE suffers less slowdown on GPUs because its encoder stack is composed of parallel hashing of $k$ encoder hash functions.
When $k\sim O(1000)$, GPU outshines CPU for such massively parallel operations.
For \textit{select}, we see only $2.1\times$ and $1.5\times$ slowdown on CPUs and GPUs, respectively.
For this \textit{select} representation, only the 3 largest embedding tables are replaced with DHE stacks.
The rest of the sparse features use table representation, leading to faster execution.
For \textit{hybrid}, we observe $11.2\times$ and $5.4\times$ slowdown on CPUs and GPUs, respectively.
\textit{Hybrid} results in longest latencies because both embedding tables and DHE stacks are executed to generate highly accurate embedding vectors.

\subsection{Accelerator Compatibility: IPU, TPU Evaluation}~\label{ssec:char_accelerators}
Next, we explore accelerator-level synergies using real modern AI accelerators: Graphcore IPUs and Google TPUs~\cite{moor2020ipu, jia2019ipu, jouppi2021tpu, jouppi2020tpu, jouppi2017tpu}.

CPU and GPU system specifications are detailed in Section \ref{ssec:method_hw}.
We evaluate TPUv3 at core-, chip-, and board-level configurations.
For chip- and board-level configurations, we employ data-parallelism for increased throughput.
We evaluate Graphcore GC200 IPU at chip-, board-, and pod-level configurations (Figure \ref{fig:ipu_parallelism}).
For a single IPU chip, table and \textit{hybrid} configurations require backup Streaming Memory (i.e., DRAM) since the model does not fit within the 900 MB on-chip SRAM.
For an IPU-M2000 board, we pipeline the model across the SRAM of the four chips.
For IPU-POD16, we employ data-parallelism.

Figure \ref{fig:accelerators} quantifies the design space trade-offs with four key observations:

\textbf{\underline{O1}: For embedding tables, TPUs achieve highest speedup due to their custom \texttt{TPUEmbedding} layers (Figure \ref{fig:accelerators} (a)).}
Each TPU core has access to 16 GB of HBM.
TPUs utilize this HBM efficiently by: 1) sharding larger tables and replicating smaller tables across TensorCore HBMs and 2) pipelining embedding lookups with TensorCore computations.
These optimizations form the basis of TPU custom \texttt{TPUEmbedding} layers.

\textbf{\underline{O2}: For DHE stacks, IPUs perform well primarily when all model parameters and inputs fit into the IPU scratchpad memory (Figure \ref{fig:accelerators} (b)).}
Each Graphcore IPU has access to 900 MB of scratchpad SRAM.
While this SRAM has no default caching abilities, when all model parameters and activations fit within this SRAM budget, IPUs rely on virtually only compute and on-chip memory accesses, leading to significant speedups.
Furthermore, the lack of off-chip DRAM access also contributes to high energy efficiency for DHE use cases.
IPU-16 achieves 16.65$\times$ performance speedup over embedding table execution on CPUs.

\textbf{\underline{O3}: GPUs offer higher energy efficiency for large embedding table-based models (Figure \ref{fig:accelerators} (bottom).}
For large embedding table execution, GPU is most energy-efficient.
This is because while TPU shows 3.12$\times$, 11.13$\times$ performance speedup for its chip- and board-level configurations, its single chip TDP is $1.8\times$ higher than that of V100's.
Additionally, the V100 is also more energy efficient than IPU for this specific use-case.
This is because a single IPU chip's SRAM scratchpad cannot hold the entire model, leading to frequent off-chip DRAM access.

\textbf{\underline{O4}: No single hardware platform is optimal for all representations and optimization objectives, motivating the need for a dynamic representation-hardware co-design solution.}
Figure \ref{fig:accelerators} shows that there is no one size fits all hardware solution across all possible embedding representations.
While TPUs accelerate embedding lookups well, when the models are small enough to fit on-chip, IPUs perform better because of more efficient on-chip memory accesses.
On the other hand, GPUs offer a competitive option from per-chip and ease-of-use standpoints (both TPUs and IPUs require lengthy one-time compilations).

The in-depth characterization based on real AI systems demonstrates the potential for performance improvement by considering heterogeneous representations and accelerators.
Thus, to best exploit these algorithmic and system level trade-offs, representation and hardware pairing must be a dynamic rather than static decision.
\section{\Name: Representation-Hardware Co-Design}~\label{sec:design}
Built upon the real system characterization results, we propose a \textbf{M}ulti-\textbf{P}ath embedding representation co-design technique for \textbf{Rec}ommendation inference, \textbf{\Name}.
\Name~maximizes throughput of correct predictions in two stages:

\textbf{Offline Stage [Section \ref{ssec:design_offline}].}
\Name~determines \textit{which embedding representation(s)} will be used and their corresponding \textit{hardware mapping strategies} (Algorithm \ref{alg:alg_offline}). Embedding representation and mapping decisions are based on system memory capacities.

\textbf{Online Stage [Section \ref{ssec:design_online}].}
\Name~considers service-level agreements (SLA), such as model accuracy and tail latency targets, of the application and runtime factors, such as input query sizes (Algorithm \ref{alg:alg_online}).
\Name~produces embedding vectors from sparse features by dynamically activating either Table, DHE, or \textit{hybrid} execution paths on available hardware platforms.

To accelerate the encoder-decoder stack in the DHE and \textit{hybrid} paths, we introduce \textbf{\CacheOne} for the encoder stack and \textbf{\CacheTwo} for the decoder stack.
\CacheOne~exploits the power law distribution of embedding access frequencies~\cite{gupta2020deeprecsys}, whereas \CacheTwo~considers value similarity of intermediate encoder stack results.
Figures \ref{fig:design} and \ref{fig:cache} provide the design overviews for \Name~and \Cache, respectively.
Next, we present the major components of \Name~in detail.

\begin{algorithm}[t!]
    \caption{\Name~Offline Stage}\label{alg:alg_offline}
    \begin{algorithmic}[1]
    \REQUIRE Embedding Representation Space $R=\{r_i\}$, Hardware Platforms $H=\{h_i\}$
    \ENSURE Optimal representation-hardware mapping strategies $S^{\star} \in \{(r_i, h_i)\}\,|\,r_i^{\star}$ is accuracy optimal.
    \STATE $S^{\star} \gets \{\}$
    \FORALL{hardware $h_i \in H$}
        \IF {$\exists r_{j,hybrid}^{\star}$ that fits on $h_i$}
        \STATE {$S^{\star} \cup (r_{j,hybrid}^{\star}, h_i)$}
        \ENDIF
        \IF {$\exists r_{j,table}$ that \textit{still} fits on $h_i$}
        \STATE {$S^{\star} \cup (r_{j,table}, h_i)$}
        \ENDIF
        \IF {$\exists r_{j,DHE}^{\star}$ that \textit{still} fits on}
        \STATE {$S^{\star} \cup (r_{j,DHE}^{\star}, h_i)$}
        \ENDIF
        \IF {$h_i$ has $\leq$ one $r_j$ mapping in $S^{\star}$}
        \STATE {$S^{\star} \cup (r_{j,DHE(compact)}, h_i)$}
        \ENDIF
    \ENDFOR
    \STATE train all representations $r_i$ found within $S^{\star}$ 
    \end{algorithmic}
\end{algorithm}
\subsection{Offline HW-Specific Representation Generation}~\label{ssec:design_offline}
\Name~factors in representation-level insights and exploration constraints to generate representation-hardware configurations (Algorithm \ref{alg:alg_offline}).

\textbf{Heterogeneous Hardware Platforms.}
\Name~considers available hardware platforms and their memory capacities for embedding access.
Traditionally, neural recommender systems are deployed exclusively on server-class CPUs and GPUs due to their large memory requirements~\cite{gupta2020architectural, hazelwood2018applied, mudigere2021zionex}.
However, with DHE's potential for compression, \Name~is able to target a wider range of hardware platforms.
For \Name, we consider each individual hardware component by their memory capacity budget and map representation(s) accordingly to maximize memory capacity utilization.

\textbf{Representation Exploration.}
Algorithm \ref{alg:alg_offline} details the steps for finding optimal representation-hardware mappings $S^{\star}$.
For each hardware component $h_i$, we first see if there exists a \textit{hybrid} embedding representation $r_{j,hybrid}^{\star}$ that is 1) under capacity budget, 2) has large \# of encoder hash functions $k$ and 3) has a decoder MLP (i.e., $d_{NN}, h$) as small as reasonably possible.
As discussed in Section \ref{sec:characterization}, we want high $k$ for better model accuracy and small decoder MLP to minimize memory footprint and FLOPs.

With a \textit{hybrid} representation that provides high accuracy, \Name~then searches for an embedding table representation $r_{j,table}$ that can be later activated to handle latency-critical situations (i.e., tight SLA latency targets).
If there is still capacity available on $h_i$, we search for a DHE representation $r_{j,DHE}^{\star}$ that has accuracy-latency trade-offs in-between the \textit{hybrid} and table configurations. We then repeat this process for all available hardware platforms.
On memory-constrained devices, we search for compact representation $r_{j,DHE(compact)}$.
Selected representations are then profiled against the expected workload at different query sizes.

With a set of representations on each hardware platform, we can selectively activate mappings during the online stage to maximize recommendation quality while hitting SLA targets and maintaining high throughput to the best of our abilities.

\begin{algorithm}[t!]
    \caption{\Name~Online Stage}\label{alg:alg_online}
    \begin{algorithmic}[1]
    \REQUIRE Representation-hardware mappings $S \in \{(r_i, h_i)\}$, Input Query $q$
    \ENSURE Selected query execution path $(r_{i}, h_{i})$
    \STATE $n,t_{SLA} \gets$ query size, SLA latency target
    \IF{$(r_{j,hybrid}, h_i)$ can process query size $n$ under $t_{SLA}$}
    \STATE {return $(r_{j,hybrid}, h_{i})$}
    \ELSIF{$(r_{j,DHE}, h_i)$ can process query size $n$ under $t_{SLA}$}
    \STATE {return $(r_{j,DHE}, h_{i})$}
    \ELSE
    \STATE {return $(r_{j,table}, h_{i})$}
    \ENDIF
    \end{algorithmic}
\end{algorithm}

\begin{table*}[t!]
\caption{Systems Configurations.}
\begin{center}\resizebox{\linewidth}{!}{
    \begin{tabular}{|c||c|c|c|c|}
    \hline
    \textbf{Machines}       & \textbf{Intel Broadwell CPU} & \textbf{NVIDIA V100 GPU} & \textbf{Graphcore IPU-M2000 (4 IPUs)} & \textbf{Graphcore IPU-POD16 (16 IPUs)} \\ \hline
    \textbf{Frequency}      & 2.2 GHz                      & 1.2 GHz                  & 1.35 GHz                              & 1.35 GHz                               \\ \hline
    \textbf{Cores}          & 12                           & 5120                     & 5888                                  & 23552                                  \\ \hline
    \textbf{Cache Sizes}    & 0.3-3-30 MB                  & 3 MB                     & 3.6 GB                                & 14.4 GB                                \\ \hline
    \textbf{DRAM Capacity}  & 264 GB                       & 32 GB HBM2        & 256 GB                                & 1024 GB                                \\ \hline
    \textbf{DRAM Bandwidth} & 76.8 GB/s                    & 900 GB/s                 & 20 GB/s                               & 80 GB/s                                \\ \hline
    \textbf{TDP}            & 105 W                        & 250 W                    & 600 W                                 & 2400 W                                 \\ \hline
    \end{tabular}}
\end{center}
\label{tbl:hw}
\end{table*}

\subsection{Online Dynamic Multi-Path Activation}~\label{ssec:design_online}
During online phase, \Name~dynamically activates representation-hardware execution paths to handle incoming queries based on query-level information (Algorithm \ref{alg:alg_online}).
Currently, scheduling is dependent on query sizes and SLA latency targets.

\textbf{Varying Query Sizes and SLA Tail Latency Targets.}
In real-world production environments, incoming queries arrive at different sizes and have to be served within application-specific SLA latency targets.
Based on prior works, recommendation workloads can have query sizes between $1-4K$ and SLA latency targets from $1-100s$ ms~\cite{gupta2020deeprecsys}.

\textbf{Maximizing Throughput of Correct Predictions.}
We activate representation execution paths based on incoming query size $n$ and SLA latency target $t_{SLA}$ (Algorithm \ref{alg:alg_online}).
If there exists a \textit{hybrid} configuration that can finish a query of size $n$ within $t_{SLA}$ without throughput degradation, that representation execution path is activated to achieve highest possible accuracy.
If no \textit{hybrid} execution path exists, we see if there is a DHE representation path for moderately improved accuracy.
If $n$ and $t_{SLA}$ are too strict, \Name~then defaults to activating the Table representation path to satisfy SLA conditions.
Ultimately, \Name~dynamically activates the path of highest recommendation quality while ensuring table-level system throughput for different SLA conditions and varying query sizes, thus maximizing the throughput of correct predictions.

\subsection{\Cache: Mitigating Latency of the Compute-Stack Path}
While DHE and \textit{hybrid} paths offer accuracy and capacity improvements, activating either path comes with significant latency overheads.
In large-scale inference serving experiments (Section \ref{sec:evaluation}), we see that these latency degradations lead to higher tail latencies.
In order to close this performance gap, we devise \textbf{\Cache}, a two-part caching optimization that exploits both access frequency and value similarity of embedding accesses (Figure \ref{fig:cache}).

\textbf{\CacheOne: Exploiting Access Frequency.}
In recommendation workloads, the access counts of \textit{power}power users and items make up a sizable portion of total accesses~\cite{wu2020developing}.
We exploit this observation by caching pre-computed embeddings of such hot, frequently accessed IDs in a cache within the encoder stage.
If we encounter a hot ID, we can directly look up the ID's pre-calculated embedding vector and skip the entire encoder-decoder stack.

\textbf{\CacheTwo: Exploiting Value Similarity.}
If a sparse feature ID does not hit in \CacheOne, it goes through the encoder stack to generate an intermediate dense vector.
This dense vector then becomes an input to the decoder MLP stack. 
As mentioned in Section~\ref{sec:characterization}, this decoder MLP stack can be costly in terms of latency.
To mitigate this latency overhead, we propose \CacheTwo~to exploit value similarity between intermediate dense vectors.
We profile the intermediate dense vectors generated from a recommendation workload's sparse IDs and construct $N$ centroids that best represent the overall distribution of possible intermediate vectors.
With these centroids, our compute becomes $k$-nearest neighbors (kNN) search between the target intermediate dense vector and $N$ centroids.
In implementation, if the vectors are normalized, finding the nearest centroid can be simplified to parallelizable dot product followed by an argmax function, thus providing speedup over computation-heavy MLP stacks.
The number of centroids $N$ is an adjustable parameter: larger $N$ gives better approximations at the cost of more compute.
\section{Methodology}~\label{sec:method}
We present the experimental methodology used for evaluating \Name~on a design space spanning hardware platforms (Section~\ref{ssec:method_hw}), recommendation use cases (Section~\ref{ssec:method_datamodel}), inference runtime characteristics (Section~\ref{ssec:method_runtime}), and evaluation metrics (Section~\ref{ssec:method_metrics}).

\subsection{Hardware Systems}~\label{ssec:method_hw}
One of \Name's core features is its ability to generate optimal representation-hardware mappings for heterogeneous systems with different memory capacities.
To demonstrate this flexibility, we evaluate \Name~at three different configurations:
\begin{enumerate}
    \item \textbf{HW-1:} Single CPU-GPU node with 32 GB CPU DRAM and 32 GB GPU HBM2. Unless otherwise specified, we evaluate this configuration in Section \ref{sec:evaluation}.
    \item \textbf{HW-2:} Resource-constrained case-study with 1 GB CPU DRAM and 200 MB GPU HBM2.
    \item \textbf{HW-3:} Custom-accelerator case-study with 32 GB CPU DRAM and board-, pod-level IPU platforms.
\end{enumerate}
CPU, GPU, and IPU performance data is collected on real commodity hardware platforms (Table \ref{tbl:hw}).
For query serving experiments involving IPU platforms (Section \ref{ssec:eval_hwx}), we exclude model compilation overheads. 
We profile IPU platforms across all possible query configurations and use this profiled information to get estimated performance.

\begin{figure*}[t!]
    \centering
    \includegraphics[width=\linewidth]{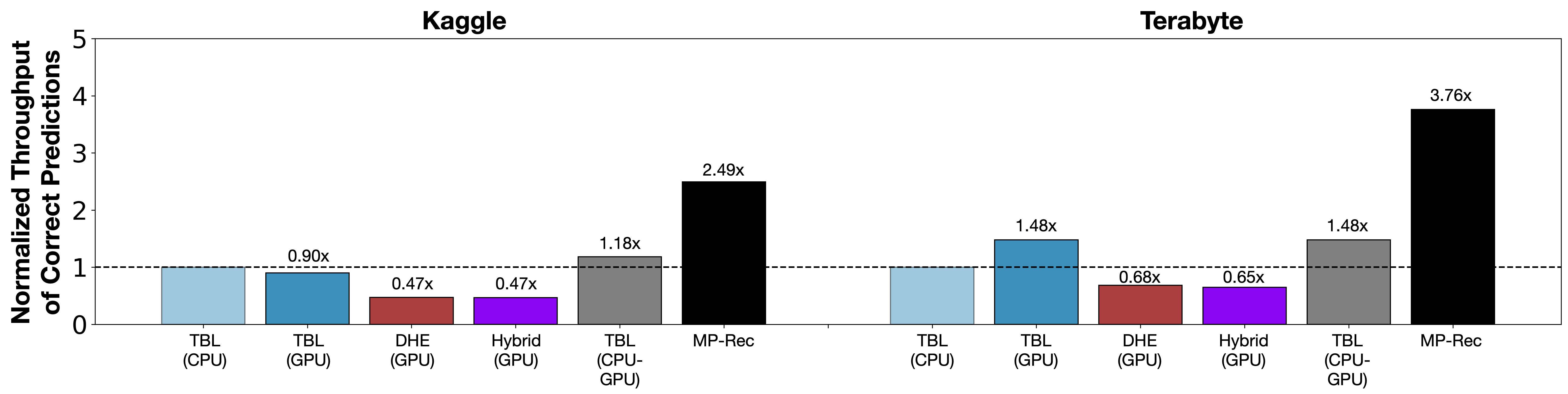}
    \caption{\textit{Throughput of Correct Predictions} for serving 10K queries in \textit{Kaggle} and \textit{Terabyte} use-cases, respectively.
    With \Cache, \Name~improves upon throughput of correct predictions by activating high accuracy execution paths.
    Statically deploying DHE, \textit{hybrid} representations leads to throughput degradations compared to executing embedding tables on CPUs and/or GPUs.}
    \label{fig:results_qpsacc}
\end{figure*}

\subsection{Datasets and Models}~\label{ssec:method_datamodel}
We evaluate \Name~with open source recommendation datasets Criteo Kaggle~\cite{criteokaggle} and Terabyte~\cite{criteoterabyte} trained on Meta's Deep Learning Recommendation Model (DLRM)~\cite{naumov2019dlrm}.
The MLPerf baseline model for Terabyte is $5.8\times$ larger than the baseline model for Kaggle (12.59 GB and 2.16 GB, respectively).
For each of the embedding representations covered in Section \ref{sec:background}, we replace the embedding tables of DLRM with our implementation of the target representation.
The encoder-decoder stack implementation is based on~\cite{kang2021dhe} and in PyTorch~\cite{paszke2019pytorch}.
Respective characterization baselines (i.e., accuracy, capacity, FLOPs) in Section \ref{sec:characterization} are from the default MLPerf DLRM-Kaggle and Terabyte configurations \cite{reddi2020mlperfinference}.
For IPU query serving use-cases (Section \ref{ssec:eval_hwx}), we reduce the embedding dimension of the Terabyte model's tables to fit the model onto IPU-POD16 (Table \ref{tbl:hw}).

\subsection{Inference Runtime Characteristics}~\label{ssec:method_runtime}
As shown in Figure \ref{fig:design}, \Name~dynamically serves inference queries across different runtime conditions:

\textbf{Query Sizes and Distribution.}
Representation-HW mappings perform differently based on query sizes.
Unless otherwise specified, we evaluate a generated query set of size 10K, following a lognormal distribution with an average query size of 128~\cite{gupta2020deeprecsys, gupta2021recpipe}.

\textbf{SLA Latency Targets.}
Inference query requests have to be finished under application-specific SLA latency targets.
For recommendation workloads, latency targets can range from $1-100s$ milliseconds~\cite{gupta2020deeprecsys}.
We overview results for a strict SLA scenario of 10ms (found in e-commerce platforms such as~\cite{cheng2016wnd, jouppi2017tpu, park2018deep, zhou2019dien, zhou2018din}) then demonstrate \Name~benefits at targets up to 200ms.

\textbf{High Throughput Inference.}
Inference engines ideally maintain high throughput.
However, large query execution may lead to QPS degradations.
Unless otherwise specified, we target 1000 QPS.

\subsection{Evaluation Metrics}~\label{ssec:method_metrics}
In addition to model quality in click-through rate prediction accuracy, and model capacity in bytes, we evaluate:
\begin{itemize}
    \item \textbf{$throughput_{correct\_predictions}$: Throughput of Correct Predictions}.
    For production use-cases, we care about not only the quality of individual recommendations but also how efficiently the models can be served at-scale. 
    We evaluate $\frac{Correct\,Samples}{Second}$ with:
    \begin{align*}
        & \frac{Queries}{Second} \times \frac{Samples}{Query} \times \frac{Correct\,Samples}{Samples} \\
        &= QPS \times Query\,Size \times Model\,Accuracy
    \end{align*}
    \item \textbf{SLA Latency Violations.}
    Meeting SLA is crucial for recommendation use cases. Thus, we also evaluate the effectiveness of \Name~in reducing SLA violations.
\end{itemize}
\section{Evaluation Results and Analysis}~\label{sec:evaluation}
In this section, we show how \Name~improves upon various static representation-hardware deployment choices -- in both throughput of correct predictions and accuracy -- by dynamically switching between representations on heterogeneous hardware.
We first evaluate \Name~-- with \Cache~enabled -- on the HW-1 design point.
Then, we cover production systems evaluation and consider resource-constrained (HW-2) and custom-accelerator (HW-3) case studies.
Next, we perform sensitivity studies on both query size distributions and SLA latency targets and explore query-splitting across heterogeneous hardware as an additional optimization.
After that, we explore \Name's dynamic switching mechanism and \Cache's two-stage structure.
Finally, we quantify how \Name~reduces SLA violations for constant throughput use-cases and present analytical scaling implications on large-scale training systems.

\subsection{\Name~Performance Overview}
\Name~achieves the highest model accuracy among all the embedding representations on both Kaggle and Terabyte datasets by using more accurate representations like DHE and \textit{hybrid} (Section \ref{ssec:eval_insights} -- \textbf{Insight 1}).
For Kaggle and Terabyte use-cases, \Name~conditionally improves achievable model accuracy by 0.19\% and 0.22\%, respectively (Table \ref{tbl:acc}).

Despite their accuracy benefits, DHE and \textit{hybrid} execution paths exhibit long latencies from their orders of magnitude higher FLOPs.
Thus, statically deploying these compute-based representations on fixed-hardware platforms leads to throughput degradations.
\Name~avoids these performance degradations by dynamically switching execution paths at the representation- and HW-level granularities (Section \ref{ssec:eval_insights} -- \textbf{Insights 2, 3}).
Furthermore, \Cache~reduces the latency of DHE and \textit{hybrid} encoder-decoder stacks, making these representations more viable for activation (Section \ref{ssec:eval_insights} -- \textbf{Insight 4}).
\Name~optimizes \newMetric~by using these factors to improve accuracy while maintaining performance.
\Name~improves \newMetric~by $2.49\times$ and $3.76\times$ on Kaggle and Terabyte, respectively (Figure \ref{fig:results_qpsacc}).
We further break down improvements in \newMetric~in Figure \ref{fig:metrics_breakdown}.

\begin{figure}[t!]
    \centering
    \includegraphics[width=\linewidth]{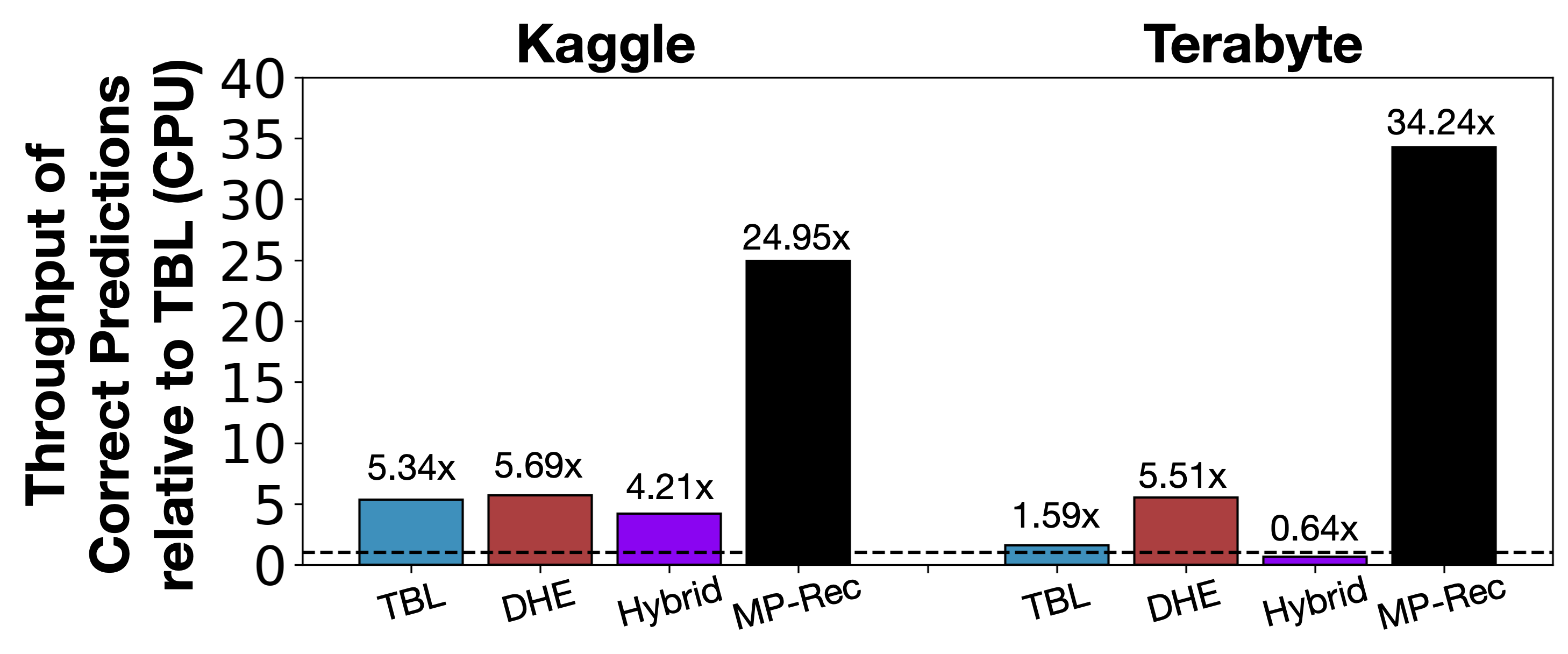}
    \caption{\textit{IPU Query Serving}: If model fits on IPUs and IPUs are able to handle dynamic query sizes, there are potential speedups across different representations.}
    \label{fig:hw_3}
\end{figure}

To achieve these benefits, \Name~stores multiple representation execution paths on each hardware platform.
This leads to increased memory footprint compared to statically using a single representation (Table \ref{tbl:results_capacity}).
We demonstrate \Name's ability to target memory-constrained and accelerator-based design points in Table \ref{tbl:hw_2} (\textbf{Insight 5}) and Figure \ref{fig:hw_3} (\textbf{Insight 6}), respectively.

\begin{table}[t!]
\caption{\textit{Achievable model accuracies} of optimal representation-hardware mappings for \textit{Kaggle} and \textit{Terabyte}, respectively. By using DHE or Hybrid, \Name~increases achievable model accuracies over table baselines.}
\begin{center}\resizebox{0.95\linewidth}{!}{
    \begin{tabular}{|c||c|c|c|c|}
    \hline
    \textbf{}                & \textbf{\begin{tabular}[c]{@{}c@{}}Table\\ (Baseline)\end{tabular}} & \textbf{DHE} & \textbf{Hybrid} & \textbf{\Name} \\ \hline
    \textbf{Kaggle}   & 78.79\%                                                             & 78.94\%      & 78.98\%         & 78.98\%         \\ \hline
    \textbf{Terabyte} & 80.81\%                                                             & 80.99\%      & 81.03\%         & 81.03\%         \\ \hline
    \end{tabular}}
\end{center}
\label{tbl:acc}
\end{table}

\begin{table}[t!]
    \caption{\textit{Memory footprints} for HW-1 on \textit{Kaggle} and \textit{Terabyte}, respectively.
    \Name~incurs greater memory footprint than static deployment choices since it stores multiple representations on each HW platform.}
    \begin{center}\resizebox{0.95\linewidth}{!}{
        \begin{tabular}{|c||c|c|c|c|}
            \hline
            \textbf{}                & \textbf{\begin{tabular}[c]{@{}c@{}}Table\\ (Baseline)\end{tabular}} & \textbf{DHE} & \textbf{Hybrid} & \textbf{\Name} \\ \hline
            \textbf{Kaggle}   & 2.16 GB        & 126 MB       & 2.29 GB         & 4.58 GB        \\ \hline
            \textbf{Terabyte} & 12.58 GB       & 123 MB       & 12.70 GB        & 25.41 GB       \\ \hline
        \end{tabular}}
    \end{center}
    \label{tbl:results_capacity}
\end{table}

\textbf{Production Use-Case Evaluation.}
We implement and evaluate \Name~using recommendation tasks in a production setting where our baseline is an internal table-based recommendation model.
We either replace the embedding tables of this baseline model with DHE stacks or augment the tables for \textit{hybrid} representations.
First, we observe a noticeable model compression ratio when replacing embedding tables with DHE stacks.
Next, \textit{hybrid} configurations achieve 0.014\% improvement in model accuracy.
Finally, utilizing DHE stacks introduces flops that incur a throughput degradation of $23.59\%$.

\subsection{Evaluation Result Insights for \Name}~\label{ssec:eval_insights}
We begin by evaluating \Name~at \textbf{HW-1} (as specified in Section \ref{ssec:method_hw}) on both the Kaggle and Terabyte use-cases.
Evaluation is done on 10K queries with targets of 1000 QPS and 10ms SLA latency target.
We highlight the key \Name~features from evaluation results:

\textbf{\underline{Insight 1}:
\Name~improves achievable recommendation quality by including carefully-tuned DHE and \textit{hybrid} execution paths.}
During its online phase, \Name~dynamically activates DHE and \textit{hybrid} paths when there is no expected latency-bounded throughput degradation.
Thus, during the execution of an entire query set, \Name~ conditionally matches higher model accuracies of DHE, \textit{hybrid} representations (Table \ref{tbl:acc}).

\textbf{\underline{Insight 2}:
\Name~mitigates the throughput degradation of DHE and \textit{hybrid} representations by conditionally activating more accurate representation(s).}
Throughput of table-only configurations on either CPUs or GPUs (blue) is higher than that of DHE-, \textit{hybrid}-only configurations (crimson, violet) (Figure \ref{fig:results_qpsacc}).
On Kaggle, using exclusively DHE or \textit{hybrid} on GPUs for their accuracy benefits degrades \newMetric~by $62.8\%$ and $63.3\%$, respectively -- compared to exclusively using embedding tables on CPUs.
This throughput degradation comes from the increased FLOPs of DHE encoder-decoder stacks.
So, even though DHE and \textit{hybrid} execution paths offer higher prediction accuracies for each \textit{individual query}, \newMetric~still drops significantly from worse system performance.
\Name~is able to recover these \newMetric degradations by selectively activating DHE, \textit{hybrid} paths based on incoming query characteristics.
Namely, \Name~schedules queries onto DHE-, \textit{hybrid}-paths when there are no expected latency-bounded throughput degradations.

\begin{figure}[t!]
    \centering
    \includegraphics[width=\linewidth]{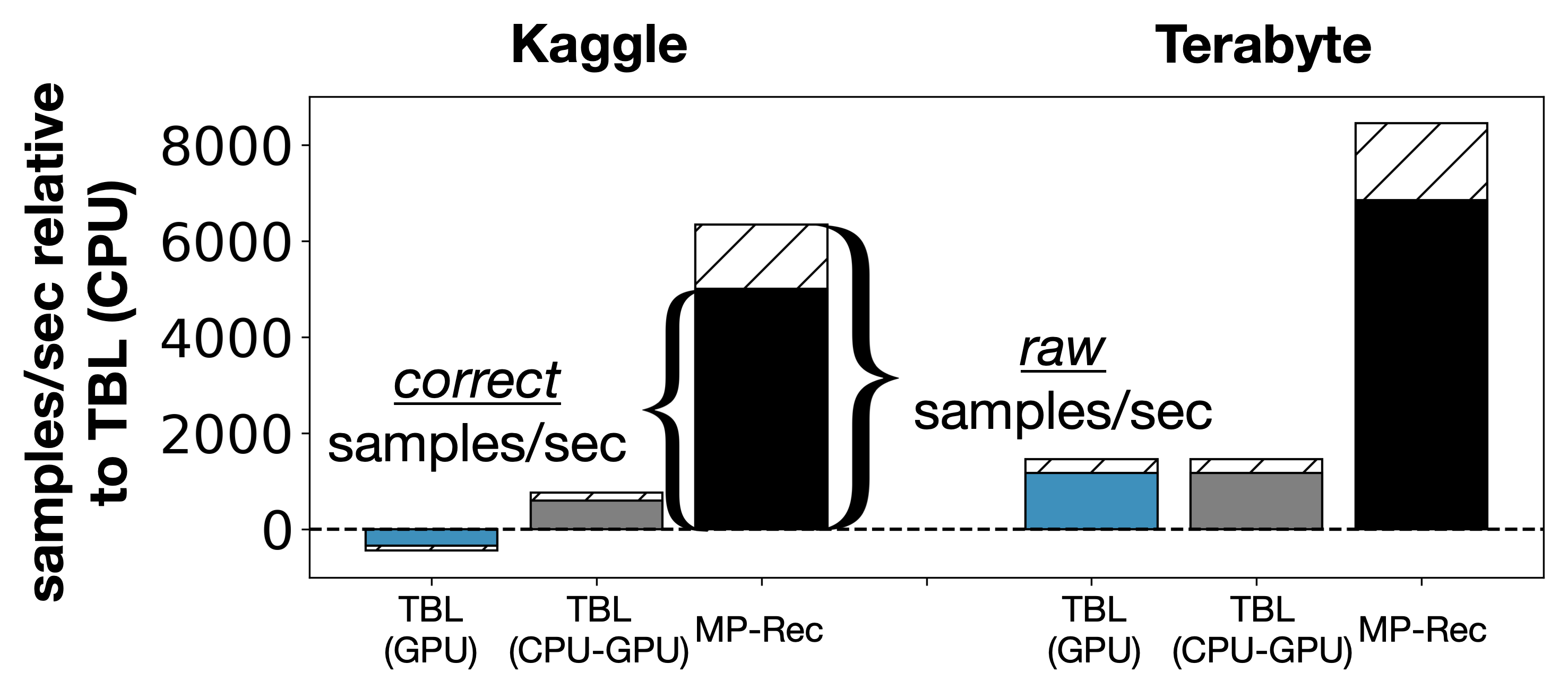}
    \caption{
    Changes in raw throughput (hatched white bars) and \textit{throughput of correct predictions} (colored bars) for \textit{Kaggle} and \textit{Terabyte} use-cases, respectively.}
    \label{fig:metrics_breakdown}
\end{figure}

\begin{table}[t!]
\caption{Achievable model accuracy, normalized throughput, and memory footprint for design point \textit{HW-2}.}
\begin{center}\resizebox{\linewidth}{!}{
\begin{tabular}{|c||c|c|c|}
\hline
                                                             & \textbf{\begin{tabular}[c]{@{}c@{}}Achievable\\ Accuracy\end{tabular}} & \textbf{\begin{tabular}[c]{@{}c@{}}Normalized Throughput\\ of Correct Predictions\end{tabular}} & \textbf{\begin{tabular}[c]{@{}c@{}}Memory\\ Capacity\end{tabular}} \\ \hline
\textbf{\begin{tabular}[c]{@{}c@{}}TBL\\ (CPU)\end{tabular}} & 78.721\%                                                                & 1.00$\times$                                                                                                 & 542 MB                                                             \\ \hline
\textbf{\begin{tabular}[c]{@{}c@{}}DHE\\ (GPU)\end{tabular}} & 78.936\%                                                               & 0.43$\times$                                                                                                  & 123 MB                                                             \\ \hline \hline
\textbf{\Name}                                               & 78.936\%                                                               & 2.26$\times$                                                                                                 & \begin{tabular}[c]{@{}c@{}}CPU: 665 MB\\ GPU: 123 MB\end{tabular}  \\ \hline
\end{tabular}}
\end{center}
\label{tbl:hw_2}
\end{table}

\textbf{\underline{Insight 3}:
\Name~improves the throughput performance of embedding table baselines by dynamically switching at the hardware platform granularity.}
Depending on use-case (i.e., base model, query statistics, and latency constraints), optimal execution paths vary by hardware platform for a particular representation.
We demonstrate this with an additional baseline where CPU-GPU switching is enabled for table-only representation (gray bars in Figure \ref{fig:results_qpsacc}).
For example, for Kaggle, purely switching at the CPU-GPU granularity achieves 18\% performance improvement over CPU-only execution (Figure \ref{fig:results_qpsacc} (left)).
However, for Terabyte, CPU-GPU switching does not enable further speedups since throughput of CPU execution, at best, matches that of GPU execution (Figure \ref{fig:results_qpsacc} (right)).
In either case, enabling CPU-GPU switching has a lower-bound performance of optimal static deployment configuration.
The reason behind this is that, for throughput, CPU execution is favored when queries are small and model complexity is relatively low (i.e., Kaggle base model).
In these scenarios, overheads for GPU-based model execution (e.g., data loading) are less amortized.

\textbf{\underline{Insight 4}:
\Cache~increases throughput of correct predictions by decreasing the latency of highly accurate representations, namely, DHE and \textit{hybrid}.}
\Cache~reduces the long latency of executing DHE, \textit{hybrid} encoder-decoder stacks.
This allows these compute-based representations to be viable for more query serving opportunities.
Without \Cache, \Name~will only switch onto long-latency execution paths when there are no expected throughput degradations.
Thus, for scenarios like large queries -- where table-based execution would have completed under strict latency targets -- \Cache~enables switching onto DHE and/or \textit{hybrid} execution paths.
\Cache~enables \Name~to improve system throughput and quality of recommendations served hand in hand.
We provide further breakdown in Figure \ref{fig:metrics_breakdown}.

\begin{figure}[t!]
    \centering
    \includegraphics[width=\linewidth]{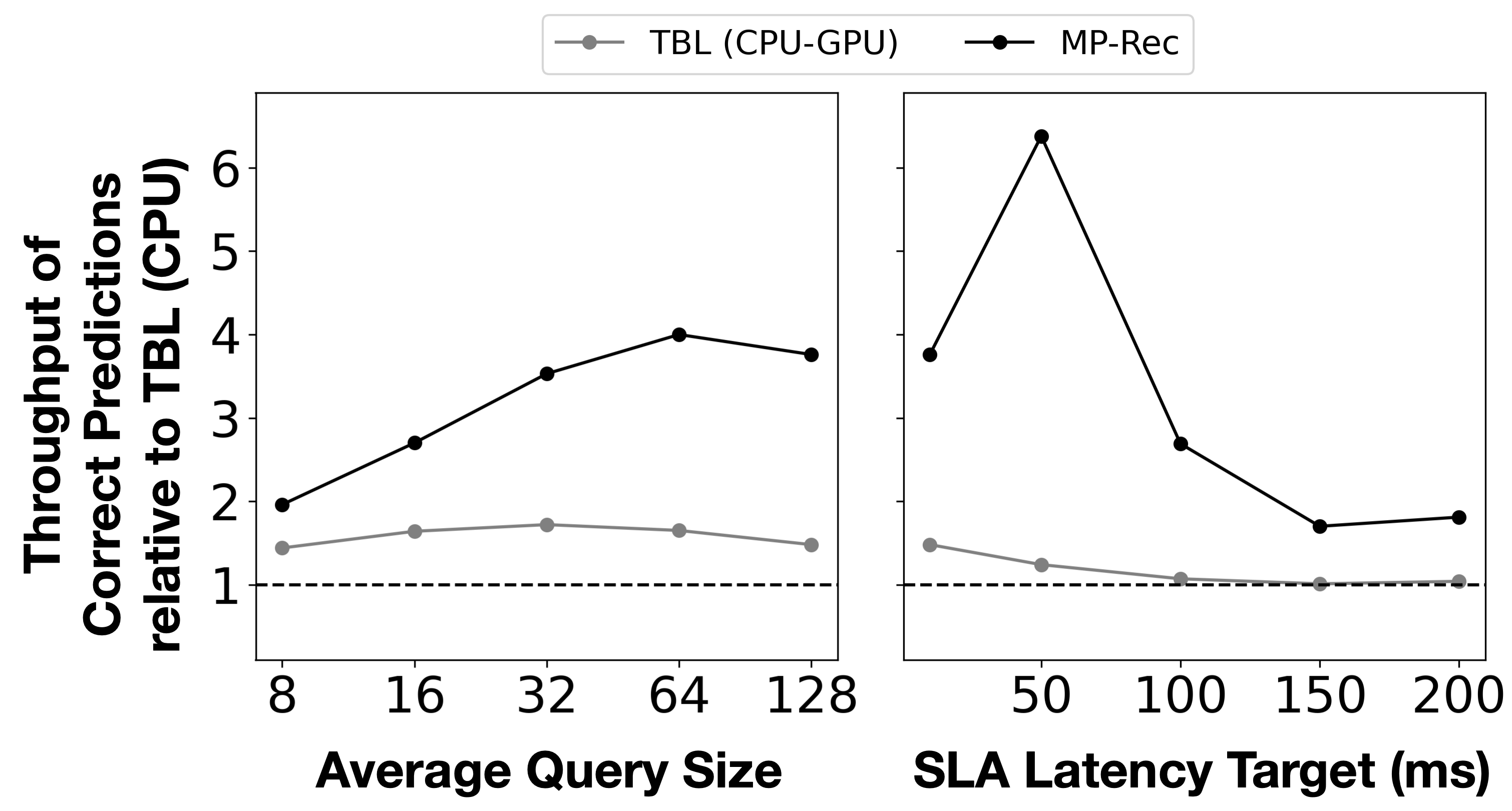}
    \caption{Sensitivity studies for query size and SLA latency target. 
    Default settings assume average query size 128 and SLA target 10 ms.
    Results shown for Terabyte use-case.}
    \label{fig:sensitivity}
\end{figure}

\subsection{Additional Heterogeneous Hardware Case Studies}~\label{ssec:eval_hwx}
In addition to evaluating ~\Name~on a large-capacity CPU-GPU system (i.e., HW-1), we expand our analysis to memory-constrained and accelerator-enabled case-studies.

\textbf{\underline{Insight 5}:
\Name~finds optimal representation-HW mappings on constrained HW design points.}
We introduce design point \textbf{HW-2} with constrained memory capacities (Section \ref{ssec:method_hw}).
As seen in Table \ref{tbl:hw_2}, \Name~utilizes HW-2's memory capacity budgets with both DHE and TBL execution paths.
By doing so, \Name~matches optimal accuracy given by DHE (Table \ref{tbl:acc})) and, at the same time, achieves higher throughput of CPU embedding table execution (Table \ref{tbl:hw_2} (left)).

\textbf{\underline{Insight 6}:
IPUs can offer potential speedups in heterogeneous hardware platforms -- given sufficiently large multi-node configurations and further software support.}
We evaluate an IPU-POD16 in our query serving experiment for both Kaggle and Terabyte (Figure \ref{fig:hw_3}).
We choose a pod-level configuration -- as opposed to chip- and board-level configurations -- since the Terabyte model is on the order of 10 GBs and accessing backup DRAM significantly hampers performance (see Section \ref{ssec:char_accelerators}).
IPU's ability to first fit entire DHE stacks on-chip then leverage data parallelism across multiple nodes enables large potential speedups on DHE and \Name~configurations.
Table and \textit{hybrid} configurations for Terabyte offer less speedup from the lack of data parallelism -- each of the 16 IPU nodes dedicate its on-chip SRAM to storing unique shards of the model's parameters.
We assume that the IPU is able to handle incoming queries of different sizes.
In practice, changes in input shapes require lengthy (i.e., $\sim30$ minutes) re-compilations.

\begin{figure}[t!]
    \centering
    \includegraphics[width=\linewidth]{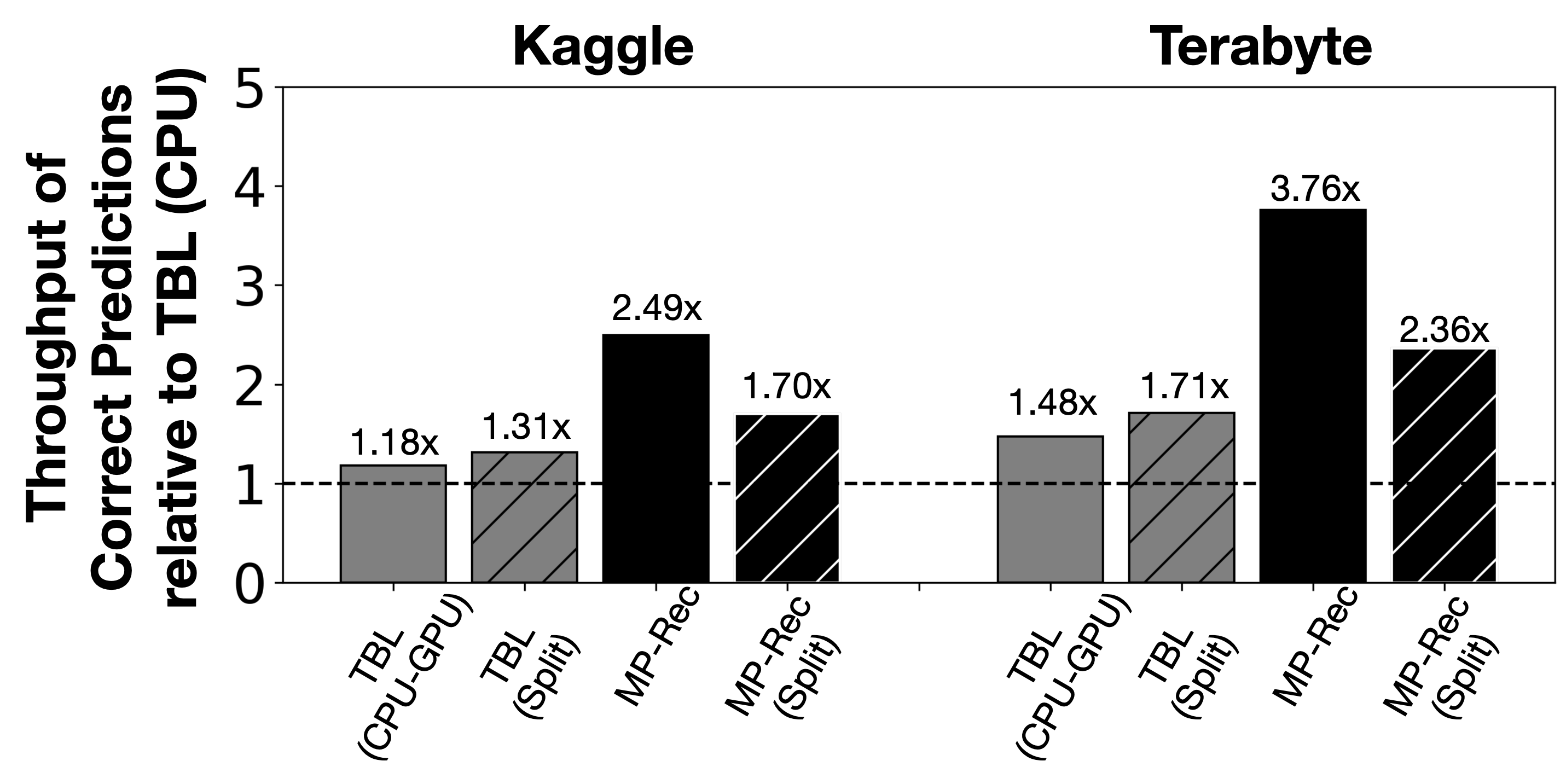}
    \caption{When incorporating DHE and \textit{hybrid}, query splitting is sub-optimal without careful tuning of split ratios. Baseline is embedding table-CPU execution.}
    \label{fig:query_splitting}
\end{figure}

\begin{figure}[t!]
    \centering
    \includegraphics[width=\linewidth]{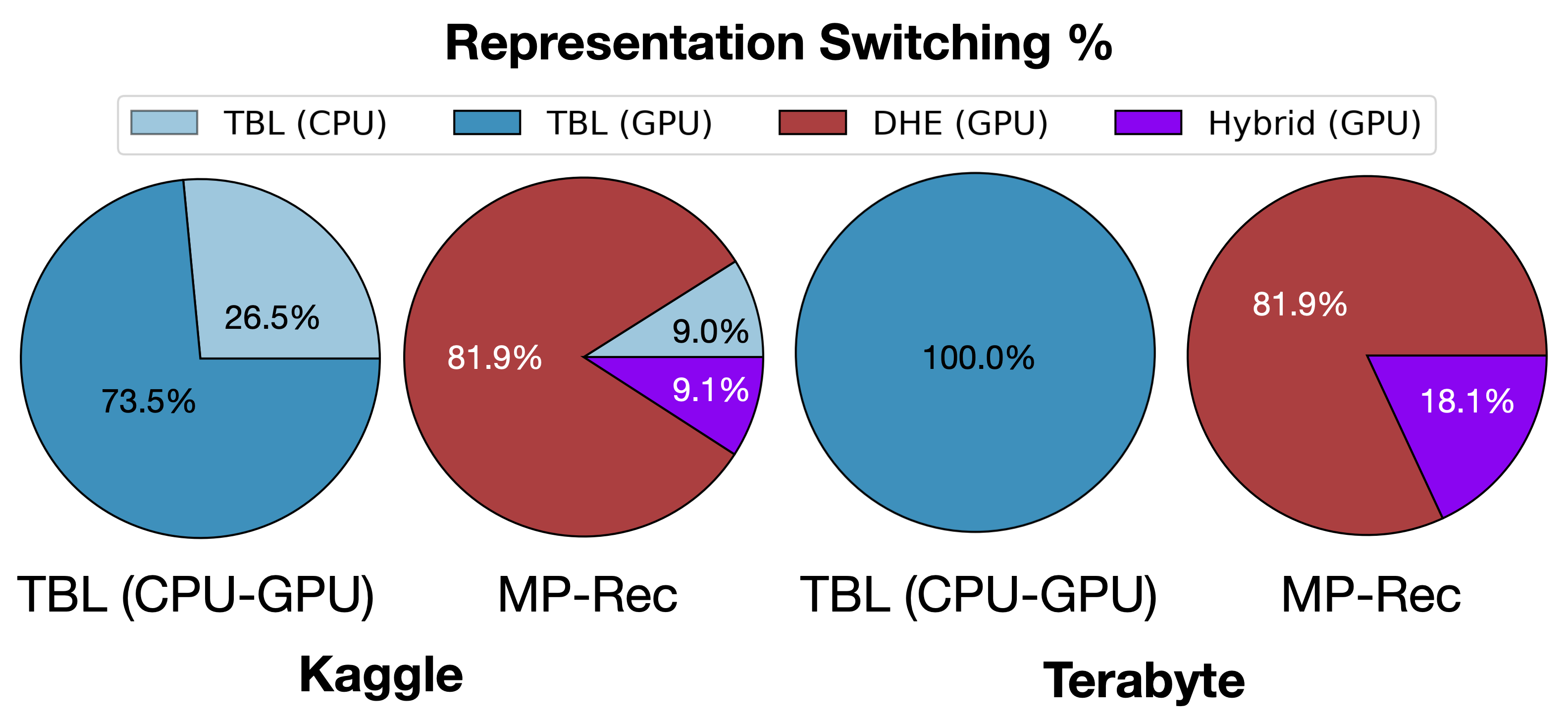}
    \caption{Switching Breakdown. \Name~enables execution of compute-based representations.}
    \label{fig:switching_breakdown}
\end{figure}

\subsection{Sensitivity Studies}
Figure \ref{fig:sensitivity} showcases sensitivity studies over the dimensions of query size and SLA latency target.
When varying average query size, we maintain a log-normal distribution for the query set (Section \ref{ssec:method_runtime}).
Figure \ref{fig:sensitivity} (left) shows that both table CPU-GPU switching and \Name~show more improvement as the average query size increases.
This is because the larger the query sizes, the more GPU/accelerator-offloading opportunities.
Figure \ref{fig:sensitivity} (right) shows that, as SLA latency target increases, speedup reduces.
This is because when latency target budget for query execution is so high (e.g., 200ms), even CPU-embedding table baseline can achieve high throughput execution.

\subsection{Additional Optimization: Query Splitting}
In order to better exploit heterogeneous hardware platforms, one potential optimization that can be applied on top of this study is query splitting. 
In theory, splitting a query for a given representation across both CPU and GPU can better utilize available resources and result in lower query load per hardware platform.
In Figure \ref{fig:query_splitting}, we explore this by evenly splitting each query for a representation across both CPU and GPU.
We observe that for embedding table configurations, query splitting is better than the CPU-GPU switching baseline.
However, for \Name, where compute-intensive representations like DHE and \textit{hybrid} are available, even query splitting is detrimental to performance.
This is because for embedding table execution, query splitting results in smaller queries that CPUs are effective for.
However, query splitting for compute-intensive representations forces CPU execution, which is extremely ineffective (see Figure \ref{fig:latency}).

\subsection{Dynamic Multi-Path Activation}

\begin{figure}[t!]
    \centering
    \includegraphics[width=\linewidth]{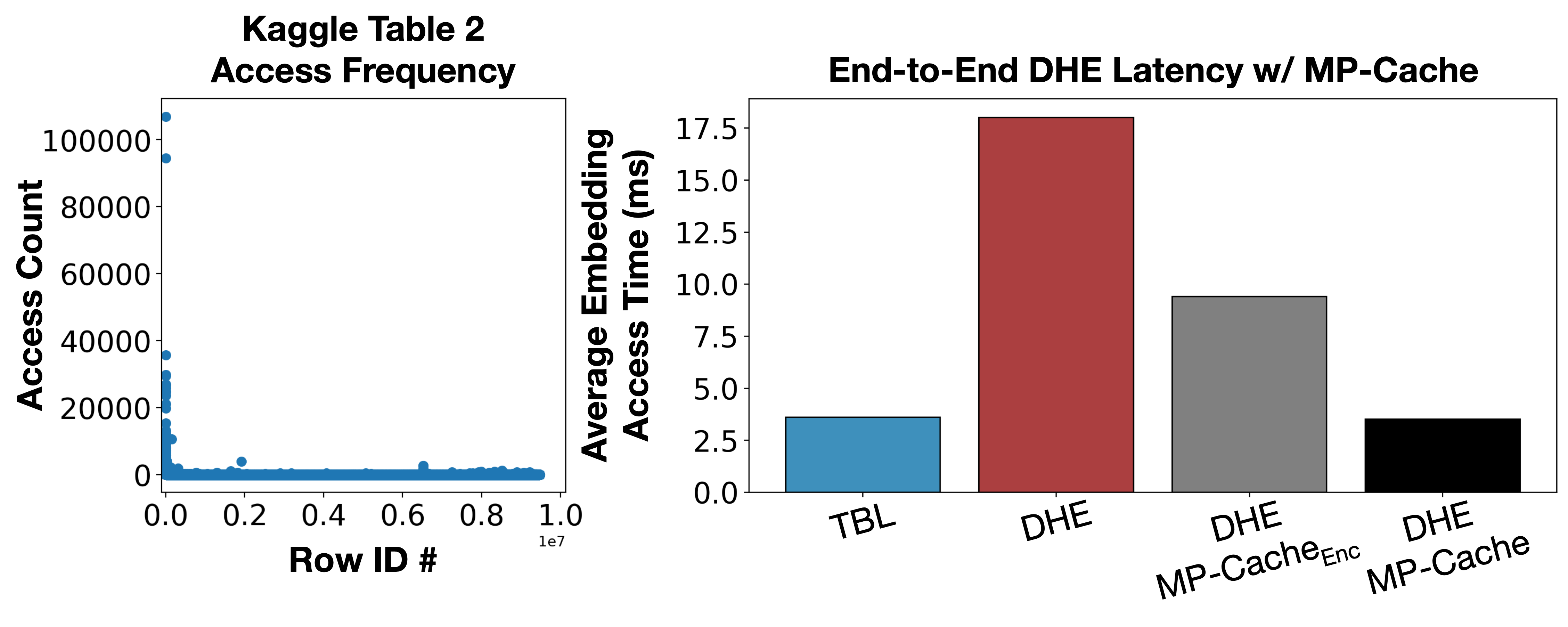}
    \caption{In recommendation workloads, ID access frequencies follow power law distributions.
    With both \CacheOne~and \CacheTwo, \Cache~closes the performance gap between encoder-decoder stacks and embedding tables.}
    \label{fig:cache_results}
\end{figure}

During the offline phase, each representation -- Table, DHE, and \textit{hybrid} -- are mapped onto both CPU and/or GPU platforms.
During the online phase, \Name~switches between representation execution paths given different input query sizes and application SLA target. 
For example, when input query is small (i.e., $\mathcal{O}(10)$), we activate table-CPU execution for tight SLA targets and \textit{hybrid}/DHE-GPU for medium SLA targets.
For large query sizes (i.e., $\mathcal{O}(100)$), table execution swaps onto GPU platform and the \textit{hybrid}/DHE-GPU path is only activated if there are no expected throughput degradations.
\textit{For each query, we activate the path of highest recommendation accuracy if it won't lead to performance degradation.
Otherwise, we activate embedding table paths to ensure throughput and latency targets are met.}

Figure \ref{fig:switching_breakdown} presents representation switching breakdown of table (CPU-GPU switching) and \Name~for both Kaggle and Terabyte.
For Kaggle, we see that TBL (CPU) is always present since the execution time of small queries on Kaggle is too fast for the GPU offloading overhead be effectively amortized.
For Terabyte, we see that TBL (GPU) is always preferable compared to TBL (CPU). This contributes to the equal performance of the TBL(GPU) and TBL(CPU-GPU) configurations, as what Figure \ref{fig:results_qpsacc} shows.

\subsection{\Cache~Result Analysis}
Figure \ref{fig:cache_results} shows how \Cache~ exploits (a) access frequency and (b) value similarity.
The access frequency opportunities come from power law distribution of recommendation workloads.
Figure \ref{fig:cache_results} (a) depicts the access distribution for Criteo Kaggle.
When we analyze the access counts of the largest sparse feature (Embedding table 2 comes with 10M entries and 3M total accesses), we confirm that hot row IDs have 10K+ access counts while others are barely accessed more than once, if at all.

\CacheOne~statically exploits the access frequency locality pattern to speed up the encoder-decoder stack.
With only 2KB dedicated to \CacheOne, we see $1.57\times$ performance improvement over using the entire encoder-decoder stack.
With a 2MB cache, the performance improvement becomes $1.92\times$.
To further close the latency gap between encoder-decoder stacks and embedding tables ($\sim5\times$ difference), \CacheTwo~converts the compute-heavy MLP in decoder stacks to kNN search.
When all vectors are normalized, kNN search becomes parallelizable dot product, leading to further speedup.
Figure \ref{fig:cache_results} (right) shows that \Cache~achieves comparable performance level for DHE as embedding table access.

\begin{figure}[t!]
    \centering
    \includegraphics[width=\linewidth]{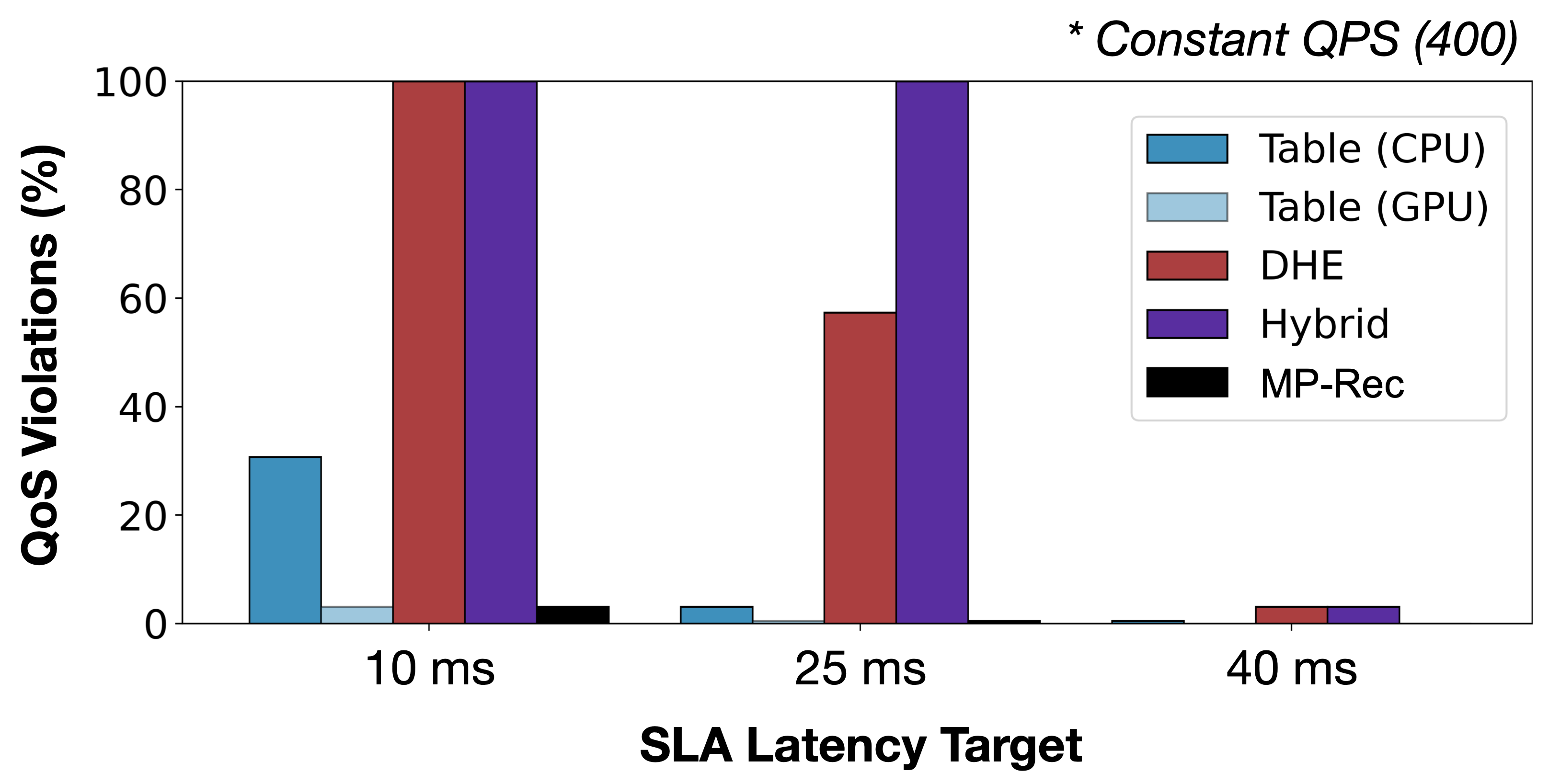}
    \caption{At constant throughput, \Name~reduces SLA latency target violations by dynamically switching to suitable representation-hardware execution paths.
    } 
    \label{fig:sla_violation}
\end{figure}

\subsection{Reducing SLA Violations}
\Name~reduces SLA target latency violations at constant throughput scenario (Figure \ref{fig:sla_violation}).
When we statically deploy a representation, SLA latency violations occur at constant throughput when the input query is too large to finish under latency target. At an SLA latency target of 10 ms, statically deploying embedding tables on CPUs will lead to 30.73\% of queries violating SLA latency target.
Without \Name~or \Cache, statically deploying DHE or \textit{hybrid} at 400 QPS will lead to 100\% SLA violation.

With \Name, we dynamically switch to representations that help us meet target SLA latency target.
Figure \ref{fig:sla_violation} shows that across a whole range of SLA latency targets, \Name~reduces percentage of queries violating SLA latency targets.
Compared to embedding table-CPU execution, \Name~observes 3.14\% SLA latency violations (27.59\% improvement) at 10ms latency target.

\begin{figure}[t!]
    \centering
    \includegraphics[width=0.95\linewidth]{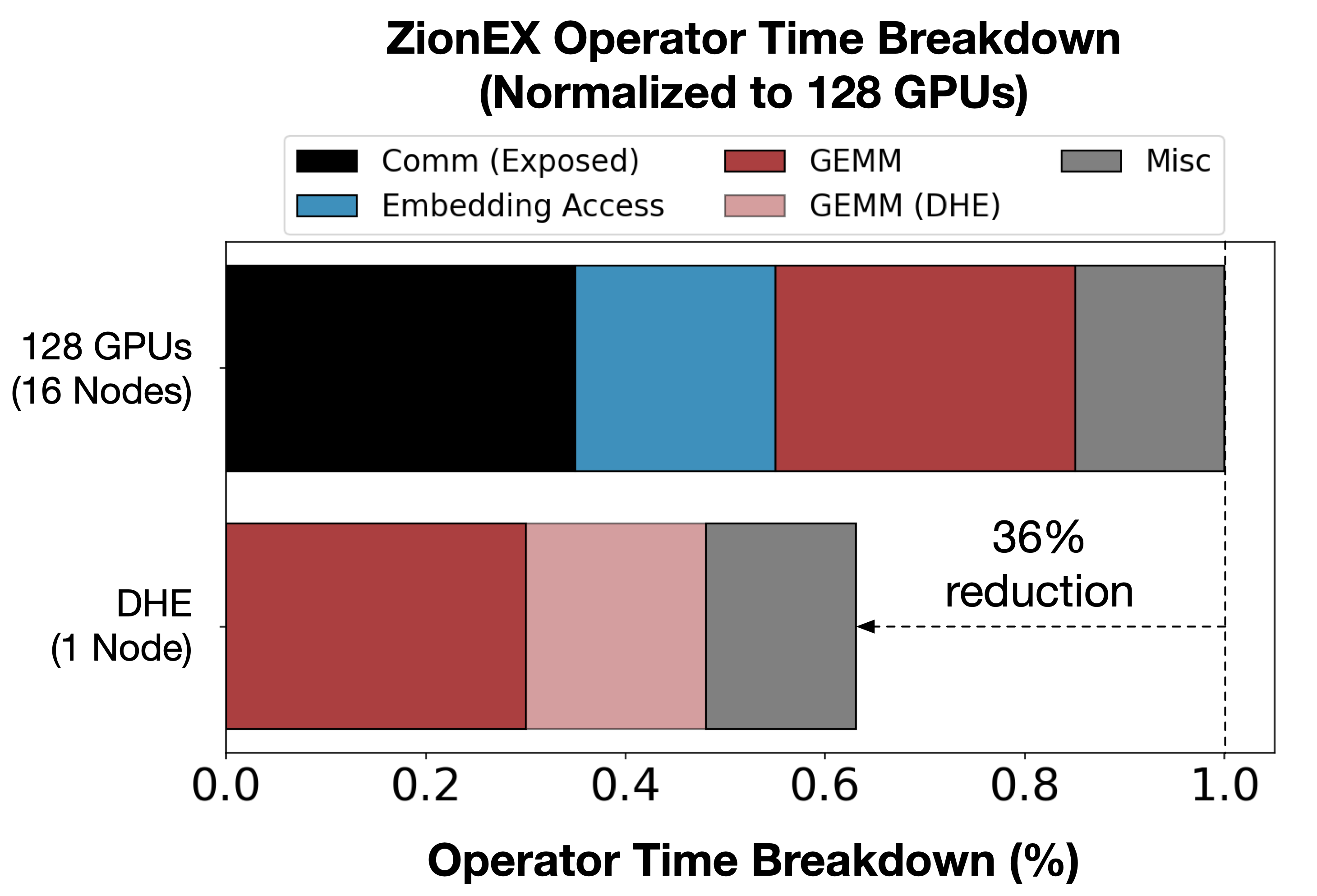}
    \caption{Through model \textit{compression}, DHE allows multi-node recommendation models to be run on a single node, reducing communication overheads.}
    \label{fig:capacity}
\end{figure}

\subsection{Scaling Analysis for Multi-Node Systems}
Production recommendation models have terabyte-scale embedding tables, leading to the requirement of multi-node hardware systems for both inference and training tasks~\cite{adnan2021high, gupta2021training, huang2021hierarchical, lui2021capacity,zhao2020baidu_dist_inf, zhao2019baidu_aibox}.
To distribute such model across multiple nodes, embedding tables have to be sharded.
During execution, communication collectives, such as \texttt{All-to-All} and \texttt{AllReduce}, are used to gather embedding-table lookup and data-parallel MLP results from different compute nodes.
These collectives contribute to inter-node communication time~\cite{sethi2022recshard}, which can be costly from a system performance perspective.
For large-scale recommendation training systems, such as ZionEX~\cite{mudigere2021zionex}, exposed inter-node communication contributes to nearly 40\% of the total model training time.

DHE can reduce the memory capacity requirement of the Terabyte benchmark by $334\times$ (Figure~\ref{fig:accuracy}).
With this compression, \Name~can potentially enable larger-size, industry-scale recommendation models for single-node systems, mitigating the exposed multi-node communication time at the cost of additional DHE-specific computation (Figure \ref{fig:capacity}).
Based on our analytical model, for a 128-GPU ZionEX system, the total execution time can be reduced by 36\% by replacing embedding tables with DHE,  thus eliminating inter-node communication~\cite{mudigere2021zionex}.
\section{Related Work}~\label{sec:related_work}
State-of-the-art neural recommender systems use embedding tables for their embedding representation, resulting in substantial memory capacity requirements~\cite{cheng2016wnd, naumov2019dlrm, zhao2019mtwnd, zhou2019dien, zhou2018din}.
Recently proposed compute-based representations reduce these memory capacity constraints at the cost of increasing FLOPs~\cite{kang2021dhe, yin2021ttrec}.
These works have explored point solutions for different embedding representations.
In this paper, we explore the interaction between embedding representations and other algorithmic and systems metrics of interest.

Earlier works overview breakdown of recommendation workloads across server-class CPU and GPU systems~\cite{acun2020understanding, gupta2020deeprecsys, gupta2020architectural, hsia2020cross, zhao2020gpuparamserver}.
Additionally, given the importance of neural recommendation, there are many proposals for custom hardware accelerators~\cite{anderson2021fbinference, gupta2021recpipe, hwang2020centaur, jiang2021microrec, ke2022axdimm, kwon2020tensor}.
In particular, many of these accelerators are DRAM-~\cite{ke2020recnmp,kwon2019tensordimm, park2021trim} and SSD-level~\cite{wan2021flashembedding, wilkening2021recssd} modifications aimed at improving embedding access time.

Other works that take advantage of heterogeneous hardware platforms include FleetRec~\cite{jiang2021fleetrec} and Hercules~\cite{ke2022hercules}.
FleetRec demonstrates that, for a fixed recommendation model, operator-level splitting is faster than CPU-only execution.
In contrast, we switch between heterogeneous embedding representations, resulting in accuracy improvements that do not exist in fixed-model approaches.
If we apply both operator-level (i.e., FleetRec) and representation-level (i.e., \Name) parallelisms for a recommendation workload, we could see improvements stemming from improved embedding table execution efficiency.
For appropriate query sizes, \textit{hybrid} execution will be activated more often due to speedups in its table stack.
However, speedup would ultimately be limited as \textit{hybrid} execution bottlenecks lie in its DHE stacks (Figure \ref{fig:latency}).
On the other hand, Hercules identifies optimal heterogeneous server architectures and system resource mappings -- especially in the context of diurnal cycles of recommendation workloads.

Other embedding table optimizations usually fall under one of two objectives: \textit{compression} or \textit{faster access}.
TT-Rec~\cite{yin2021ttrec}, ROBE~\cite{desai2022robe}, and~\cite{ginart2019mixeddim,shi2019qr} leverage matrix factorization and other related weight sharing/reduction techniques to reduce overall memory footprint of embedding tables.
Other works such as cDLRM~\cite{balasubramanian2021cdlrm}, Bandana~\cite{eisenman2018bandana}, RecShard~\cite{sethi2022recshard}, DreamShard~\cite{zha2022dreamshard}, AutoShard~\cite{zha2022autoshard}, FlexShard~\cite{sethi2023flexshard}, and Kraken~\cite{xie2020kraken} leverage access frequency information to make embedding access more hardware-efficient.
\section{Conclusion}~\label{sec:conclusion}
We explore alternative embedding representations for deep learning recommendation, where embedding vectors can be generated from embedding table lookups and/or encoder-decoder stacks.  
This is fundamentally different from prior work, where user-item relationships are learned using a single feature representation.
To maximize throughput of correct predictions while meeting tail latency requirements, we propose a new representation-system co-design approach for real-time inference, \Name. 
Depending on memory capacities of AI inference systems, \Name~selects unique embedding representations, forming multiple embedding execution paths for recommendation inference. 
At runtime, depending on input query sizes and application-dependent performance targets, \Name~activates embedding path(s) to jointly maximize model quality and throughput. 
Using the open-source MLPerf-DLRM with Kaggle and Terabyte datasets, \Name~achieves higher throughput of correct predictions and model quality at the same time while meeting application-specific tail latency requirements. 
\section{Acknowledgements}~\label{sec:acknowledgements}
We would like to thank Jay Li and Sherman Wong for the countless discussions and invaluable feedback on recommendation model architecture design and experimentation across Meta's workloads.
This collaboration not only helped refine representation designs but also enabled initial production evaluation. 

\bibliographystyle{plain}
\bibliography{refs}

 \appendix
 \section{Artifact Appendix}

\subsection{Abstract}

This artifact package includes CPU- and GPU-compatible PyTorch-based implementations of proposed deep learning recommendation architectures (i.e., embedding representations), range of relevant hyperparameters, and custom TPU-, IPU-compatible  implementations.
The base implementation is compatible with PyTorch-compliant CPUs and GPUs while TPU-, IPU-implementations require access to cloud-hosted TPUs/IPUs and their associated PyTorch branches (i.e., PyTorch/XLA and poptorch).
Inference experiments and characterization require at least a single CPU/GPU node (TPU-IPU benchmarking scripts support single- and multi-node execution) while design space exploration involving accuracy evaluation is best executed with large-scale GPU clusters due to the large number of GPU training jobs.
Open-sourcing these implementations of proposed embedding representations will allow other researchers to not only characterize and deploy these algorithmic innovations on their own choice of systems but also develop novel embedding processing techniques of their own.

\subsection{Artifact check-list (meta-information)}

{\small
\begin{itemize}
  \item {\bf Algorithm: }embedding representations; Deep Learning Recommendation Model (DLRM); recommendation systems
  \item {\bf Compilation: }(for TPU) XLA; (for IPU) PopART
  \item {\bf Model: }Deep Learning Recommendation Model (DLRM) -- publicly available at \url{https://github.com/facebookresearch/dlrm} (2 $\sim$ 12GB, depending on model architecture configuration); other model architecture modifications in artifact packages
  \item {\bf Data set: }Criteo Kaggle/Terabyte benchmarks -- publicly available via MLCommons MLPerf benchmarks; artifact also provides instructions on how to synthetically generate Kaggle/Terabyte-like input data for characterization purposes
  \item {\bf Run-time environment: }Linux
  \item {\bf Hardware: }(base benchmarks) CPU, GPU; (additional benchmarks) TPU (1$\sim$8 core configurations), IPU (1$\sim$16 chip configurations)
  \item {\bf Run-time state: }yes -- embedding access patterns may affect measured system performance
  \item {\bf Execution: }sole-user; profiling
  \item {\bf Metrics: }execution time; (commented out) operator breakdown; model accuracy
  \item {\bf Output: }profiled timing information; (commented out) operator breakdowns; training accuracy
  \item {\bf Experiments: }base experiments print out overall timing information (operator breakdown commented out)
  \item {\bf How much disk space required (approximately)?: }(without Criteo Kaggle/Terabyte datasets) 50 GB
  \item {\bf How much time is needed to prepare workflow (approximately)?: }(without setting up Criteo Kaggle/Terabyte datasets and on CPU-GPUs) 20 min; (TPU-IPU benchmarks) associated software stack setups (see official PyTorch-TPU/IPU tutorials) require 1+ hr
  \item {\bf How much time is needed to complete experiments (approximately)?: }inference runs take minutes to reach steady state; training runs take 18$\sim$24 hours per run -- fully exploring design space requires 100s of training runs.
  \item {\bf Publicly available?: }Yes
  \item {\bf Code licenses (if publicly available)?: }MIT License
  \item {\bf Data licenses (if publicly available)?: }Criteo Terabyte License (\url{https://ailab.criteo.com/criteo-1tb-click-logs-dataset/})
  \item {\bf Workflow framework used?: }(training runs) slurm
  \item {\bf Archived (provide DOI)?: }Yes; \url{https://github.com/samhsia/MP-Rec-AE}
\end{itemize}
}

\subsection{Description}

\subsubsection{How to access}

Clone repository from \url{https://github.com/samhsia/MP-Rec-AE}

\subsubsection{Hardware dependencies}

Any CPU/GPU should be able to run the reference recommendation workloads. 
As described in the paper, we experimented on server-class (Xeon) Intel CPUs and NVIDIA (CUDA-enabled) GPUs.
Any CPU/GPU platform that is compatible with PyTorch (see \textbf{Software dependencies} subsection -- also duplicated in $\texttt{dlrm\_mprec/requirements.txt}$) will work.
TPUs and IPUs available via cloud are compatible with the custom TPU and IPU implementations that leverage modules within their associated software packages.

\subsubsection{Software dependencies}

PyTorch; PyTorch/XLA; poptorch.
Required PyTorch packages:
\begin{itemize}
    \item future
    \item numpy
    \item onnx
    \item pydot
    \item torch
    \item torchviz
    \item scikit-learn
    \item tqdm
    \item torchrec
    \item torchx
    \item primesieve
\end{itemize}

\subsubsection{Data sets}

Criteo Kaggle/Terabyte.
For instructions on how to generate data in the shape of Criteo Kaggle and Terabyte datasets, see $\texttt{dlrm\_mprec/configurations.txt}$.
We also provide a script to download Criteo Kaggle in $\texttt{dlrm\_mprec/download\_kaggle.sh}$.

\subsubsection{Models}

Base DLRM and variations: DHE,\textit{hybrid}, and \textit{select}.
See paper for more description and chracterization on these variations and $\texttt{dlrm\_mprec/configurations.txt}$ for sample commands for running each of these variations on both CPUs and GPUs.

\subsection{Installation}

Use $\texttt{pip}$, $\texttt{conda}$, or your choice of Python package manager to install requirements listed above in \textbf{Software dependencies} subsection (also in $\texttt{dlrm\_mprec/requirements.txt}$).

\subsection{Evaluation and expected results}

$\texttt{python dlrm\_s\_pytorch.py --mini-batch-size=2}$ \\$\texttt{--data-size=6}$ should be your first command to ensure within installation works.
If the aforementioned command works, you should try running the other embedding representations with commands listed in: $\texttt{dlrm\_mprec/configurations.txt}$.
You should see profiled timing information.

\subsection{Experiment customization}

Listed in $\texttt{dlrm\_mprec/configurations.txt}$ are DHE stack hyperparameters relevant to the characterization and discussion in the paper. We list ranges for number of encoder hash functions and decoder MLP shapes.

\end{document}